\documentclass{aa}  

\usepackage{graphicx}
\usepackage{txfonts}
\usepackage{lipsum}
\usepackage{subcaption}
\usepackage{lscape}
\usepackage{placeins}
\usepackage{amsmath}
\usepackage{enumitem}
                                
\usepackage[colorlinks=true,allcolors=blue]{hyperref}

\newcommand{\bm}{$\langle B \rangle$}
\newcommand{\bz}{$\langle B_{\rm z} \rangle$}
\newcommand{\vsini}{$v_\mathrm{e}\sin i$}
\newcommand{\kms}{km\,s$^{-1}$}
\newcommand{\orcidlink}[1]{\protect\href{https://orcid.org/#1}{\protect\includegraphics[width=8pt]{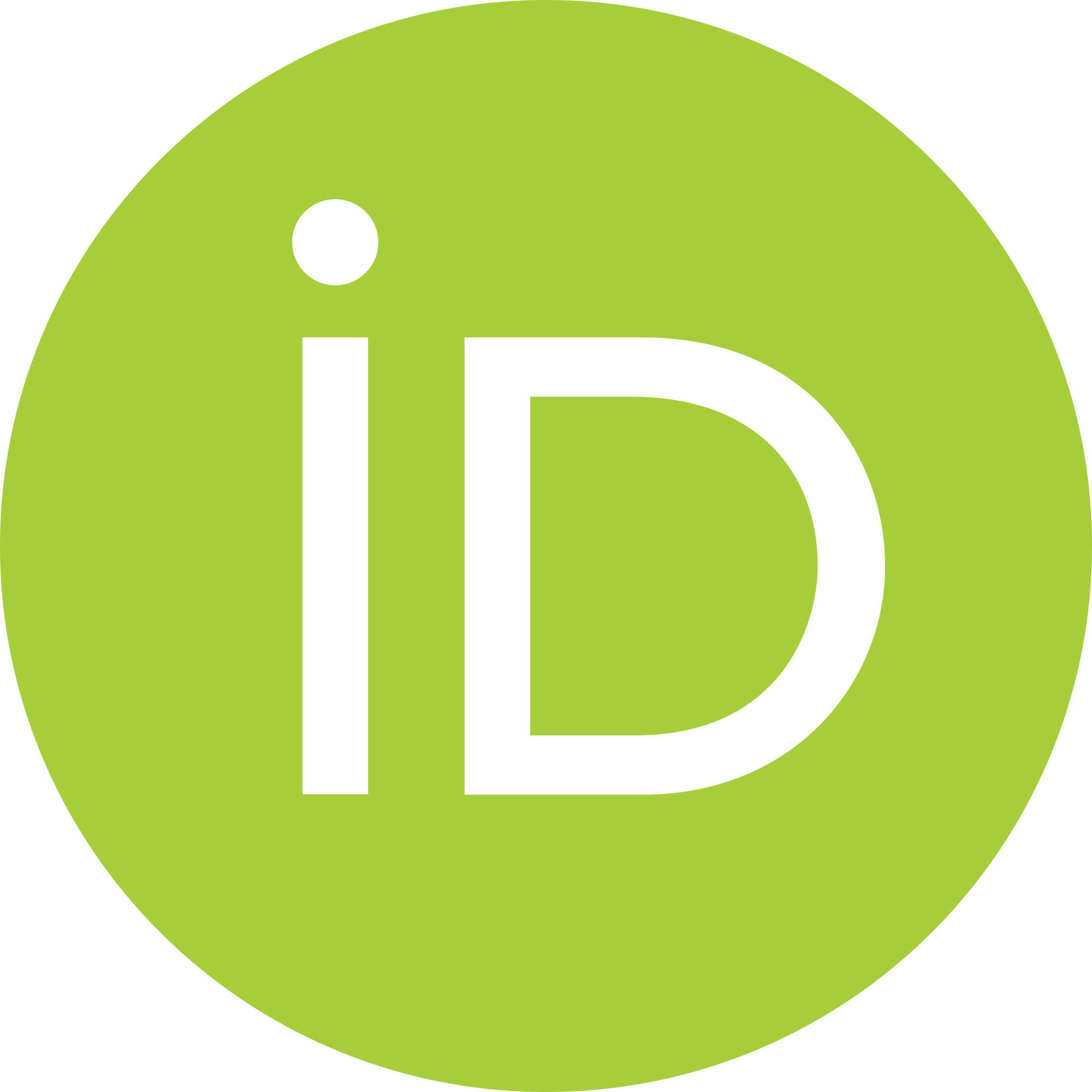}}}

\makeatletter
\let\linenumbers\relax
\let\endlinenumbers\relax
\makeatother

\begin{document}

\title{How to interpret near-infrared polarisation spectra \\ of active M dwarfs?}

\titlerunning{Interpreting M-dwarf polarisation spectra}

\author{Oleg Kochukhov\thanks{Corresponding author: \email{oleg.kochukhov@physics.uu.se}}\orcidlink{0000-0003-3061-4591}}
\authorrunning{Oleg Kochukhov} 

\institute{Department of Physics and Astronomy, Uppsala University, Box 516, SE-751 20 Uppsala, Sweden}

\date{Received 30 December 2025 / Accepted 26 January 2026}

\abstract
{Analyses of global magnetic fields in M dwarfs rely on many approximations regarding the derivation of average line profiles from spectropolarimetric data, interpreting them with analytical functions and modelling them using Zeeman Doppler imaging (ZDI). These assumptions have not been systematically tested, especially for near-infrared observations that require a more accurate treatment of the Zeeman effect.}
{We assessed the accuracy of standard treatments of average polarisation profiles in M dwarfs and their interpretation with ZDI. We focused on the filling-factor approach, which attempts to represent coexisting global and small-scale fields.}
{We performed polarised radiative transfer calculations across the near-infrared spectrum of a typical M dwarf, including atomic and molecular opacities. From these theoretical spectra, we derived mean Stokes profiles and approximated them with different line-synthesis methods. To test the recovery of global fields, we performed ZDI inversions using simulated Stokes $V$ observations for low- and high-activity cases.}
{The analytical approximation of mean polarisation profiles reproduces Stokes $I$ and $V$ only for fields up to $\sim$1 kG and fails for linear polarisation. ZDI with single-line analytical Stokes $V$ profiles is adequate for weakly magnetic M dwarfs with fields below a few hundred gauss. However, combined with the filling-factor formalism, this traditional modelling approach produces unphysical local fields and distorted global geometries for active M dwarfs with multi-kilogauss fields. These issues are mitigated using a new mapping technique based on theoretical Stokes profiles that account for both global and randomly distributed small-scale fields.}
{Our study reveals fundamental limitations of current ZDI analyses of active M dwarfs and questions the reliability of some published maps. The new computational framework proposed here enables a more accurate and physically consistent interpretation of high-resolution polarimetric spectra, representing a step towards multi-scale magnetic diagnostics of low-mass stars.}


\maketitle

\section{Introduction}
\label{sec:intro}

M-dwarf stars play a central role in modern astrophysics, largely due to their abundance in the local Galaxy \citep{kirkpatrick:2024} and the increasing number of exoplanets discovered in their habitable zones \citep[e.g.][]{dressing:2015}. Their magnetic fields govern many aspects of their behaviour, from starspot formation and high-energy flaring to angular momentum loss and rotational evolution \citep[e.g.][]{mcquillan:2013,yang:2017,see:2025}. Understanding these magnetic fields is essential for characterising the radiation environments experienced by orbiting planets, which directly affect atmospheric escape, surface conditions, and overall planetary habitability \citep{kislyakova:2017,alvarado-gomez:2022,vidotto:2023}. Moreover, the strong magnetic activity typically exhibited by M dwarfs provides a natural laboratory for studying dynamos operating in fully convective interiors, a regime that differs considerably from solar-type stars \citep{browning:2008,yadav:2015,kapyla:2021}. As interest in low-mass stars continues to grow, accurately constraining the magnetic properties of M dwarfs has become fundamental not only for stellar physics but also for assessing the likelihood of life beyond the Solar System.

Two primary observational techniques have emerged as the leading approaches for probing magnetic fields in M-dwarf atmospheres: polarimetry and Zeeman broadening analysis \citep{kochukhov:2021}. Polarimetric observations, particularly those employing Zeeman Doppler imaging (ZDI) inversions \citep{kochukhov:2016}, allow us to reconstruct large-scale magnetic field topologies by modelling rotationally modulated polarised line signatures \citep[e.g.][]{morin:2008,morin:2010,kochukhov:2019a,lehmann:2024,donati:2025}. This method is uniquely capable of revealing the vector structure of global fields but is inherently insensitive to the small-scale magnetic flux that is cancelled out in polarised signals. In contrast, Zeeman broadening methods measure magnetic intensification and line splitting in unpolarised spectra, providing an estimate of the total unsigned magnetic flux present on the stellar surface \citep[e.g.][]{shulyak:2017,reiners:2022,cristofari:2023}. While this approach captures both large- and small-scale magnetic components, it does not yield directional information and often requires complex radiative transfer modelling to disentangle magnetic effects from stellar atmospheric parameters. Together, these two methods offer complementary insights but also introduce interpretive challenges when their results diverge.

A growing body of observational work has revealed significant discrepancies between magnetic field strengths inferred from polarimetry and those derived from Zeeman broadening, underscoring the multi-scale complexity of M-dwarf magnetism. Polarimetric analyses typically recover field strengths far weaker than the total fields inferred from unpolarised line broadening, often by an order of magnitude or more \citep{reiners:2009,kochukhov:2019a,kochukhov:2021}. This disparity implies that a substantial fraction of magnetic energy resides in small-scale, mixed-polarity structures that escape detection by polarimetry but strongly impact spectral line shapes. Such hierarchical magnetic structuring is consistent with numerical models of dynamo in fully convective stars; such models predict turbulent, locally highly inhomogeneous magnetic configurations that nevertheless possess a significant global component \citep[e.g.][]{yadav:2015}. The observational mismatch between different magnetic diagnostics therefore provides important clues about the underlying dynamo processes, but it also highlights the need for a unified framework capable of reconciling measurements made at different spatial scales.

Despite its success in mapping global magnetic geometries, the polarisation technique -- especially in the context of ZDI -- relies on several assumptions that have not been rigorously validated for M-dwarf atmospheres. For instance, the method uses average polarisation profiles -- least-squares deconvolved (LSD) spectra \citep{donati:1997,kochukhov:2010a}  -- to extract polarimetric signatures from noisy observational data. The LSD profiles are derived under the weak-field approximation and are then interpreted in inversions with a simplified single-line model based on analytical solutions of the polarised radiative transfer equation \citep{morin:2008,donati:2025a}. The validity of this approach was thoroughly tested and is widely accepted for optical observations of hotter stars with weak fields \citep{kochukhov:2010a}. However, M dwarfs often host kilogauss-strength magnetic fields and are increasingly observed at near-infrared wavelengths, where the weak-field approximation and the assumption of the self-similarity of Stokes $V$ profiles are more prone to failure due to the $\lambda^2$ scaling of Zeeman splitting. Additionally, molecular opacities produce a background that hinders any direct interpretation of LSD profiles, which require an ad hoc re-normalisation before any quantitative analysis. Finally, the application of ZDI to active M dwarfs can no longer ignore the presence of very strong small-scale fields despite the fact that these field structures are cancelled out in polarisation spectra. Instead, the observed polarisation profiles bear the signature of small-scale fields through their unusual shape and excessive broadening. To model this effect, ZDI studies introduce further ad hoc modifications, such as the filling factor formalism. The latter has never been properly motivated or tested.

Without addressing these methodological limitations, the reliability of global magnetic field maps derived with ZDI -- and their comparison to fields inferred from Zeeman broadening -- remains an open question, necessitating both a critical reassessment of published results and a more rigorous validation of common modelling approaches. In this study, we present the first systematic evaluation of the LSD and ZDI frameworks as applied to M dwarfs by simulating stellar near-infrared spectra that contain representative mixtures of global and small-scale magnetic structures. Our approach begins by synthesising Stokes parameter spectra with the highest feasible realism across broad wavelength intervals, incorporating detailed polarised radiative transfer and relevant atomic and molecular opacities. We then process these simulated data using the same analysis methods routinely employed in observational studies, thereby subjecting the techniques to conditions that closely mirror real applications. This enables us to evaluate the accuracy of the widely used analytical approximations for LSD Stokes parameter profiles under spectral resolutions, wavelength coverages, and spectral line choices representative of modern M-dwarf spectropolarimetry. We subsequently apply ZDI to the simulated spectropolarimetric time series and, by comparing the reconstructed magnetic configurations with the known input fields, quantify the fidelity of the inversion algorithms, identify systematic failure modes, and examine the robustness of several key computational assumptions. This methodology provides, for the first time, a controlled environment in which to explore the strengths and shortcomings of commonly used polarisation diagnostics for M dwarfs.

The remainder of the paper is organised as follows. In Sect.~\ref{sec:methods} we describe in detail our treatment of the superposed large- and small-scale surface magnetic fields, the computation of polarised spectra that includes both atomic and molecular contributions, the construction and interpretation of LSD profiles, and the procedures used to recover global magnetic geometries with ZDI. Section~\ref{sec:results} then examines the results produced at each stage of this pipeline, emphasising where approximations begin to diverge from the more realistic synthetic data. Specifically, in Sect.~\ref{sec:local} we assess the agreement between synthetic LSD profiles and the simplified analytical line profile models widely used in the community, highlighting the conditions under which these approximations remain valid or break down. Sections~\ref{sec:zdi-weak} and \ref{sec:zdi-strong} present the results of our ZDI inversions for weakly and strongly magnetic M dwarfs, respectively, comparing standard line-modelling practices with the revised approach developed in this work and demonstrating how these choices affect the recovered magnetic topology. Finally, Sect.~\ref{sec:disc} offers concluding remarks, including a broader discussion of our findings and a critical re-evaluation of recent literature results in the context of the validation tests performed here.

\section{Methods}
\label{sec:methods}

\begin{figure*}[t!]
\centering
\includegraphics[height=8.6cm]{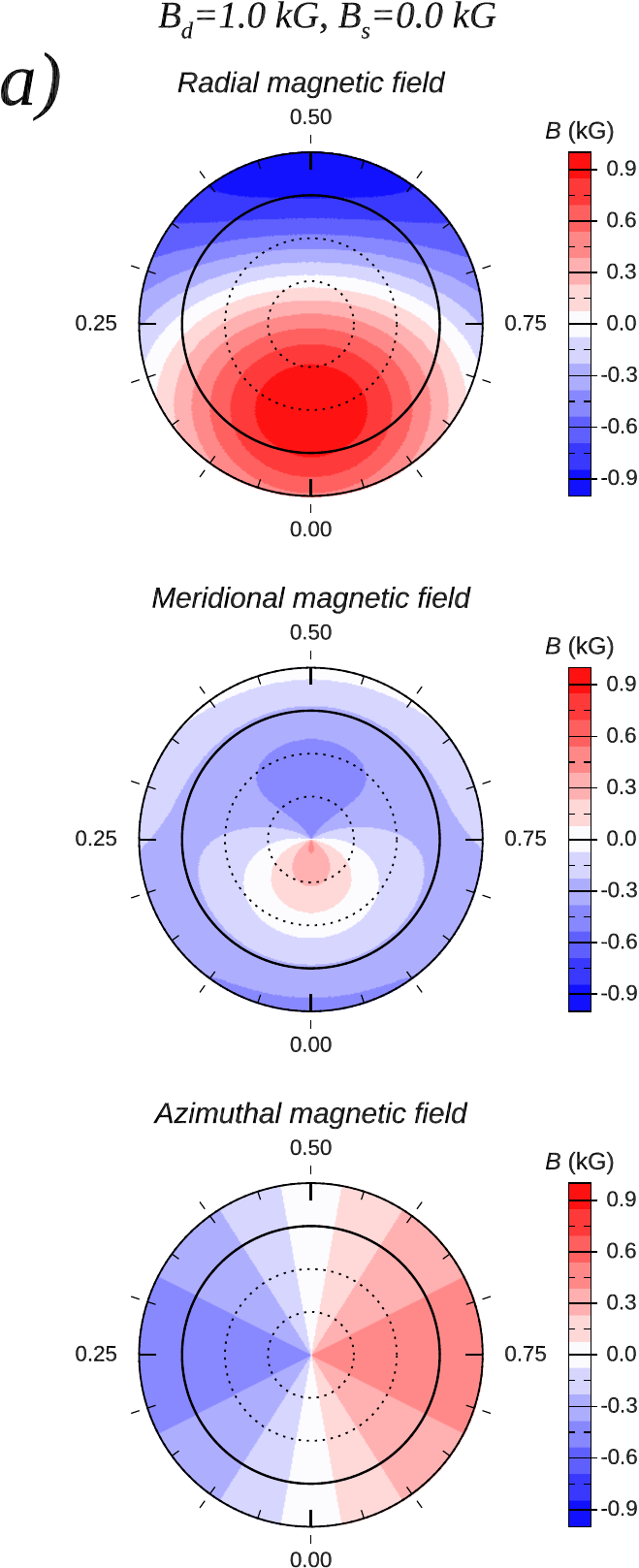}\hspace*{5mm}
\includegraphics[height=8.6cm]{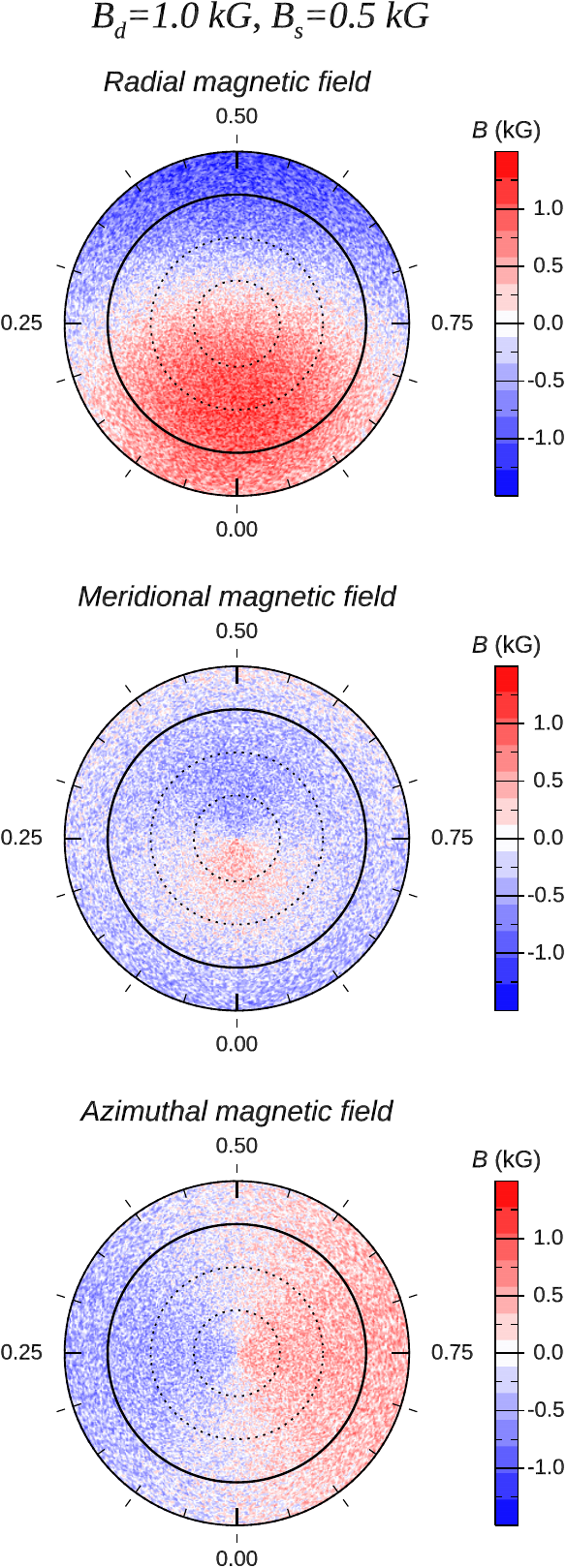}\hspace*{5mm}
\includegraphics[height=8.6cm]{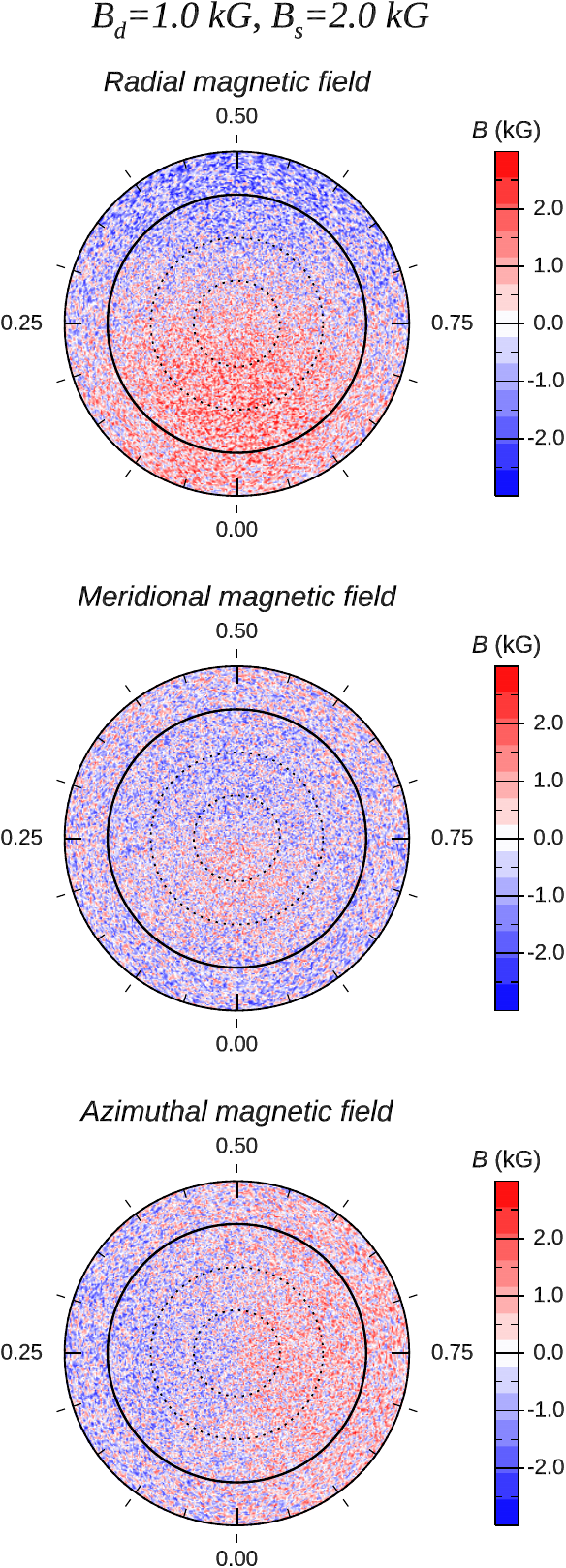}\hspace*{5mm}
\includegraphics[height=8.3cm]{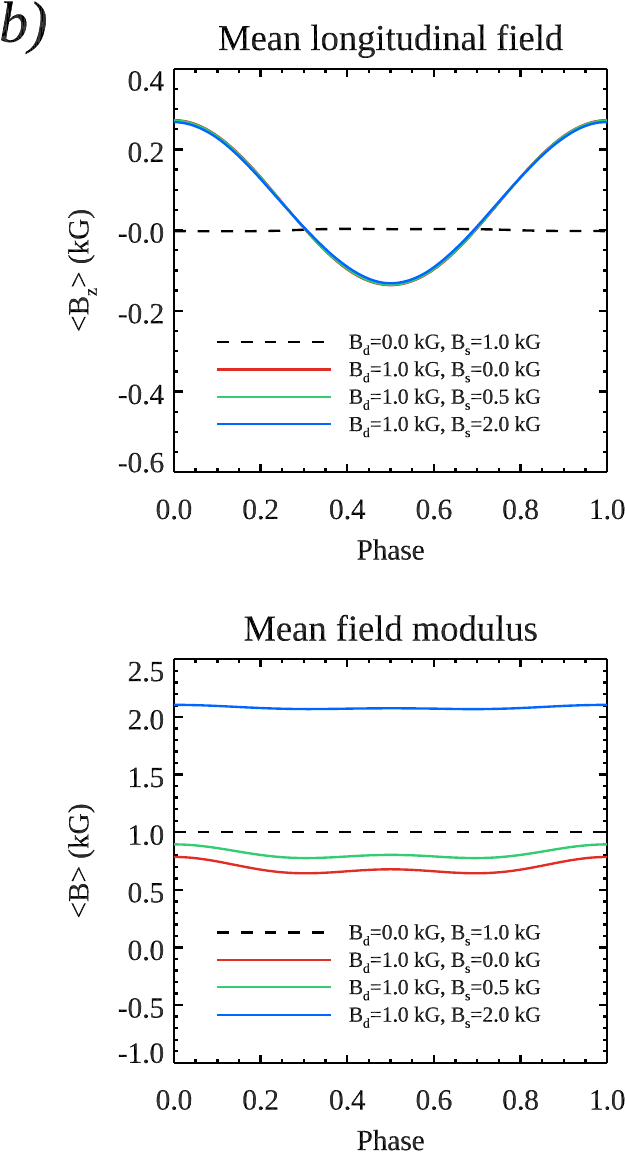}
\caption{Composite magnetic field configurations and corresponding integral magnetic observables for superpositions of global and small-scale magnetic fields. \textit{Columns in panel (a)}: Maps of the combined magnetic field for the same global dipolar geometry ($B_{\rm d}=1$~kG, $\beta=60\degr$) with an added isotropic small-scale field of varying strength (0--2~kG, indicated above each column). These plots display flattened polar projections of the radial, meridional, and azimuthal magnetic field components. In each polar plot, the thick circle marks the stellar equator, and dotted lines indicate latitude intervals of 30\degr. \textit{Panel (b)}: Mean longitudinal magnetic field, \bz\ (\textit{top}) and the mean field modulus, \bm\ (\textit{bottom}) for the three superpositions of small-scale and global fields (solid lines) as well as for the small-scale field alone (dashed lines).}
\label{fig:msup}
\end{figure*}

\subsection{Superposition of small-scale and global fields}
\label{sec:msup}

While the coexistence of a large-scale, ordered magnetic field with a much stronger background of tangled local magnetic structures is commonly discussed in studies of M-dwarf magnetism, no quantitative parametric description of such superpositions has so far been incorporated into modelling efforts. Here we represent this multi-scale magnetic topology using a simple composite model that combines a global field geometry -- assumed to be described by an oblique dipolar field $\vec{B}_{\rm d}$ -- with a uniform, isotropic small-scale field component of a single strength $B_{\rm s}$. The latter is generated by assigning random orientations to magnetic field vectors of equal magnitude, defined locally on the stellar surface. This formulation is designed to capture the dominant observational effects of multi-scale magnetic fields while remaining intentionally simple. It provides a controlled way to combine large- and small-scale magnetic contributions in spectropolarimetric modelling, without introducing many specific assumptions about unresolved surface structure.

For a stellar surface grid consisting of $N$ elements, the Cartesian components of this random field are computed as
\begin{align}
B_{\rm s}^x & =  B_{\rm s} \sin{\theta}\cos{\varphi}, \nonumber \\
B_{\rm s}^y & =  B_{\rm s} \sin{\theta}\sin{\varphi}, \\
B_{\rm s}^z & =  B_{\rm s} \cos{\theta}, \nonumber
\end{align}
where
\begin{align}
\theta &= \arccos(2 t_1 - 1), \nonumber \\
\varphi &= 2 \pi t_2,
\end{align}
with $t_1$ and $t_2$ being two independent sequences of $N$ uniformly distributed random numbers in the interval $[0,1]$. The total magnetic field at each surface element is then obtained by a vector sum of the global and small-scale components,
\begin{equation}
\vec{B}_{\rm tot} = \vec{B}_{\rm d} + \vec{B}_{\rm s}.
\end{equation}

This formulation can be straightforwardly generalised to non-dipolar global fields and can accommodate an arbitrary distribution of small-scale field strengths, as typically inferred from Zeeman broadening studies \citep[e.g.][]{shulyak:2017,reiners:2022}. In this case, a multi-component small-scale field ($\vec{B}_{\rm s}$) is constructed as a weighted sum of random field realisations with different characteristic strengths,
\begin{equation}
\vec{B}_{\rm s} = \sum_i f_i \vec{B}^{i}_{\rm s},
\end{equation}
where each $\vec{B}^{i}_{\rm s}$ represents a random field distribution of a given strength, and the weight $f_i$ denotes the fraction of the stellar surface covered by small-scale magnetic field of that strength.

The approach presented here is conceptually similar to the treatment of small-scale fields in the magnetospheric modelling of \citet{lang:2014}, where such fields were represented by a carpet of small circular spots with alternating radial polarities. However, in our formulation the small-scale field is not restricted to being purely radial, nor do we explicitly impose the condition of zero net signed magnetic flux when integrated over the stellar surface. In practice, numerical tests demonstrate that for $N\sim10^5$ the resulting $\vec{B}_{\rm s}$ exhibits negligible signed magnetic flux and makes no discernible contribution to net polarisation diagnostics, such as the mean longitudinal magnetic field \bz.

\begin{figure*}[t!]
\centering
\includegraphics[height=0.90\hsize,angle=90]{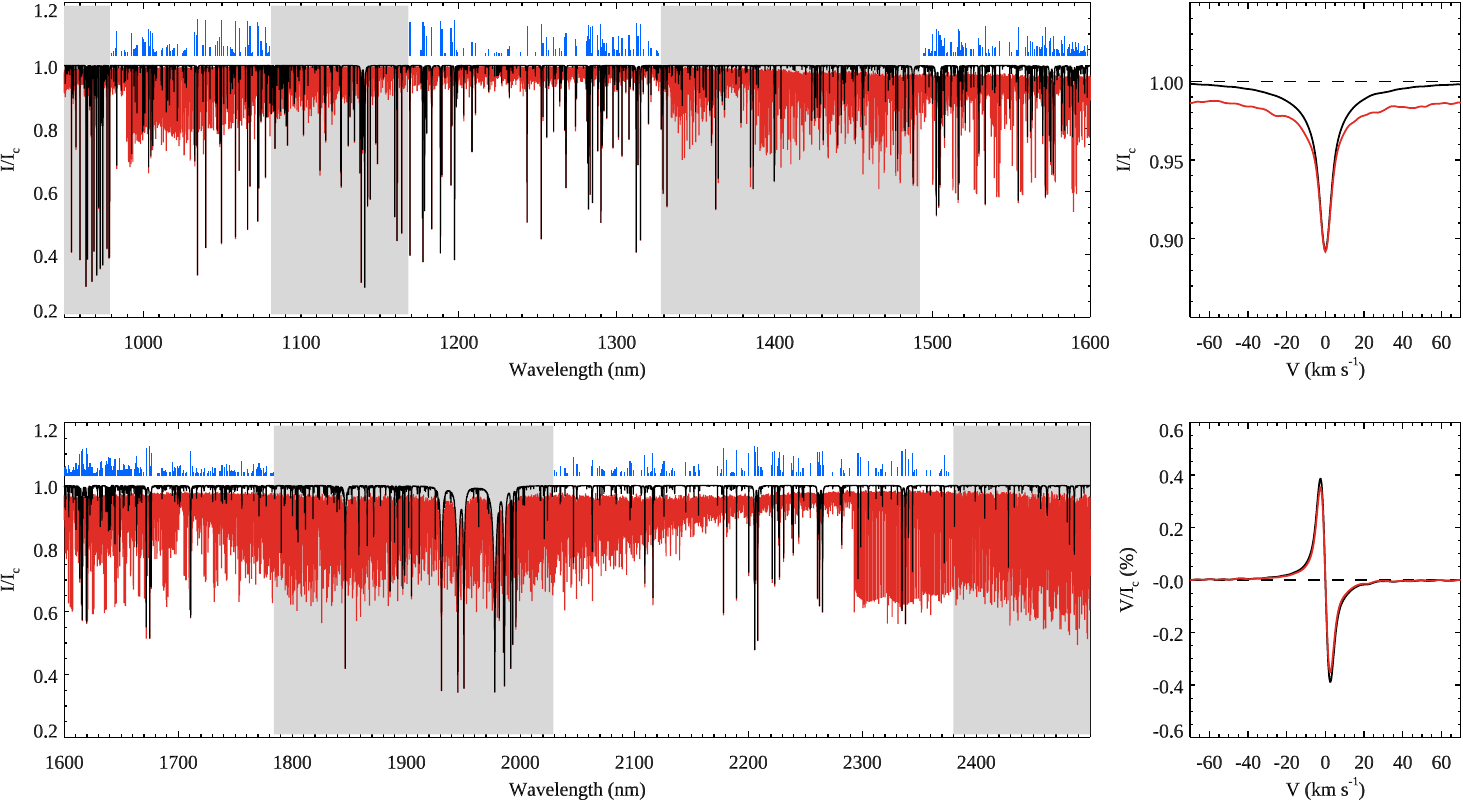}
\caption{Theoretical near-infrared M-dwarf synthetic spectra and corresponding LSD profiles. \textit{Left panels}: Disk-centre Stokes $I$ spectra computed for a model atmosphere with $T_{\rm eff}=3500$ K and $\log g = 5.0$, both with (red) and without (black) molecular line opacity. A 100~G line-of-sight magnetic field was adopted for these calculations. Blue bars above the spectra mark the positions of atomic lines included in the LSD mask, with bar lengths proportional to their central depths. Grey rectangles highlight wavelength regions excluded from the analysis because they are heavily contaminated by telluric absorption in typical ground-based observations. \textit{Right panels}: Corresponding LSD profiles: Stokes $I$ (\textit{upper}) and Stokes $V$ (\textit{lower}).}
\label{fig:LSD-mol}
\end{figure*}

Figure~\ref{fig:msup}a illustrates magnetic field structures for different superpositions of $\vec{B}_{\rm d}$ and $\vec{B}_{\rm s}$. The leftmost column shows the radial, meridional, and azimuthal field components for a purely global dipolar field with the polar strength $B_{\rm d}=1$~kG and obliquity $\beta=60\degr$. The following two columns present the composite field topology resulting from the superposition of the dipolar field and small-scale random fields with strengths of half and twice $B_{\rm d}$. The integral characteristics of these composite fields are examined in Fig.~\ref{fig:msup}b, which presents the rotational phase variation of the mean field modulus \bm\ and the mean longitudinal magnetic field \bz. These quantities are obtained through numerical surface integration of $\vec{B}_{\rm tot}$ for the cases shown in Fig.~\ref{fig:msup}a, assuming an inclination angle $i=60\degr$ and a linear limb-darkening law with coefficient $\eta=0.21$. As is evident from the \bz\ curves (top panel in Fig.~\ref{fig:msup}b), the inclusion of the small-scale field leaves longitudinal field essentially unchanged, while significantly increasing the mean field modulus (bottom panel in Fig.~\ref{fig:msup}b) -- precisely the behaviour intended with this composite multi-scale field geometry framework.

\subsection{Polarised spectrum synthesis}
\label{sec:prt}

In this study we employed the polarised radiative transfer code {\sc Synmast} to compute realistic, high-resolution local Stokes parameter spectra. The thermodynamic structure of the line-forming layers was represented by a solar-metallicity {\sc MARCS} model atmosphere \citep{gustafsson:2008} with $T_{\rm eff}=3500$~K and $\log g=5.0$, typical for a mid-M dwarf. Owing to the absence of readily available one-dimensional model atmospheres for magnetised low-mass stars, we assumed that a single atmospheric model describes the temperature and pressure stratification over the entire stellar surface, irrespective of the local magnetic field strength.

The {\sc Synmast} calculations made use of an atomic line list extracted from the {\sc VALD} database \citep{ryabchikova:2015} for the selected {\sc MARCS} atmosphere over the wavelength range 950--2500~nm. This interval matches the spectral coverage of the SpectroPolarimètre InfraRouge \citep[SPIRou;][]{donati:2020} instrument, which has been widely used in recent spectropolarimetric investigations of M-dwarf stars \citep{bellotti:2024,lehmann:2024,donati:2025a,donati:2025}. The spectra were sampled at a constant velocity step of 0.5~\kms. In total, 3247 atomic lines were included in the synthesis, with full Zeeman splitting patterns treated explicitly. Land\'e factors were taken from VALD where available or otherwise computed assuming LS coupling.

Although molecular lines dominate the spectra of M dwarfs, their response to magnetic fields is generally weak. Even the most magnetically sensitive molecular band heads, such as those of TiO and FeH, produce polarisation signals that are small compared to those of typical atomic lines \citep{afram:2019,kochukhov:2021}. Nevertheless, molecular absorption plays a role in shaping Stokes $I$ spectra by significantly depressing the continuum level and, consequently, shifting continuum in the corresponding LSD profiles. To account for this effect, we incorporated an extensive set of near-infrared molecular absorbers into the {\sc Synmast} calculations. Molecular opacities were precomputed for the adopted model atmosphere using a new, highly optimised non-magnetic spectrum synthesis code, {\sc FastSpec}, described in Appendix~\ref{sec:fs}. These opacities were derived from a combined list of $\approx$\,$135\times10^6$ molecular lines belonging to 17 isotopologues and subsequently imported into {\sc Synmast}, where they were added to the continuum opacity.

Using this setup, we computed local Stokes $IQUV$ spectra on a dense grid spanning magnetic field strengths from 0 to 10~kG, 20 limb angles, and 20 inclinations of the magnetic field vector relative to the observer’s line of sight. The magnetic field strength and orientation were assumed to be constant with depth in the stellar atmosphere. In addition, a separate grid of continuum Stokes $I$ spectra was calculated over the same wavelength range and limb-angle sampling. All calculations assumed zero micro- and macroturbulent velocities, consistent with the low convective velocities predicted by three-dimensional hydrodynamic simulations of mid- and late-M dwarfs \citep{wende:2009}. The synthetic line Stokes spectra were convolved with a Gaussian kernel with ${\rm FWHM}=4.28$~\kms, corresponding to the nominal resolving power $R=70\,000$ of SPIRou. 

Figure~\ref{fig:LSD-mol} (left panels) shows a representative example of local synthetic spectrum computed with {\sc Synmast}, both with and without the inclusion of molecular opacity from {\sc FastSpec}. The inclusion of molecular opacity results in a systematic depression of the Stokes $I$ continuum by several per cent across the entire H and K bands, even prior to the application of rotational Doppler broadening. This continuum suppression is dominated by the cumulative contribution of numerous overlapping H$_2$O absorption lines and is expected to become progressively more pronounced in cooler atmospheres and in stars with higher projected rotational velocities, where line blending is further enhanced.

The resulting database of local Stokes parameter spectra was subsequently used to synthesise disk-integrated intensity, circular, and linear polarisation spectra for prescribed magnetic field geometries, stellar inclinations, projected rotational velocities, and rotational phases. As discussed in Sect.~\ref{sec:msup}, a surface discretisation with $N\sim10^5$ elements is required to ensure complete cancellation of the isotropic, random small-scale magnetic field component. We therefore adopted a surface grid of $N=100\,114$ elements, with latitude-dependent partitioning into surface zones of approximately equal area \citep{piskunov:2002a}. For each surface element, the local Stokes spectra were obtained by linear interpolation within the precomputed grids as a function of magnetic field strength, field orientation, and limb angle. The disk-integrated spectra were then constructed by summing the contributions from all visible surface elements, weighted by their projected areas, Doppler-shifted according to stellar rotation, and scaled by the corresponding interpolated local continuum intensities. During the disk-integration step, the Stokes $Q$ and $U$ parameters were rotated into the observer’s reference frame to account for the local azimuth of the magnetic field vector, using the standard transformation relations \citep[e.g.][]{polarization:2004}.

\subsection{LSD profile calculations}
\label{sec:lsd}

All previous studies of large-scale magnetic fields in M dwarfs have relied on LSD profiles. This spectropolarimetric observable is designed to optimally combine information from hundreds to thousands of spectral lines under the assumption that individual Stokes line profiles are self-similar and differ only by a scaling factor \citep{donati:1997}. This assumption is strictly valid only in the weak-field regime ($\la$\,1--2~kG at optical wavelengths), where Zeeman splitting remains small compared to intrinsic line broadening \citep{kochukhov:2010a}. Nevertheless, when interpreted with appropriate forward modelling, LSD profiles retain significant diagnostic value even in the presence of stronger magnetic fields \citep{kochukhov:2014}. In this work, we computed LSD profiles using the {\sc iLSD} code and followed the procedures described by \citet{kochukhov:2010a}.

Recent near-infrared spectropolarimetric studies of M dwarfs differ in their criteria for selecting spectral lines used in LSD calculations. For instance, \citet{donati:2023} retained only lines deeper than 10 per cent of the unpolarised continuum, whereas \citet{fouque:2023} and \citet{bellotti:2024} included all lines with depths exceeding 3 per cent of the continuum level. In the present study, we adopted an intermediate threshold of 5 per cent, which resulted in a total of 1370 atomic lines within the analysed wavelength range.

A characteristic challenge of ground-based near-infrared observations is the presence of strong telluric absorption over substantial parts of the spectrum. Although advanced telluric correction techniques are routinely applied \citep[e.g.][]{artigau:2014a,smette:2015}, some wavelength intervals remain effectively unusable due to near-saturated telluric absorption. To account for this limitation, we excluded the wavelength regions [950, 979], [1081, 1169], [1328, 1492], [1784, 2029], and [2380, 2500]~nm, following the prescription of \citet{bellotti:2024}. After applying these exclusions, the number of atomic lines available for LSD analysis was reduced to 785. The left panels of Fig.~\ref{fig:LSD-mol} illustrate the excluded wavelength intervals together with the positions of the lines retained in the LSD mask. The resulting line list is characterised by the following mean parameters: wavelength $\lambda_0=1630.6$~nm, effective Land\'e factor $z_0=1.19$, and central depth $d_0=0.177$. These values were used to normalise the LSD line mask weights, as described by \citet{kochukhov:2010a}.

This LSD framework was employed in two distinct parts of our analysis. First, using the disk-integrated spectra obtained in Sect.~\ref{sec:prt}, we generated simulated observations by resampling the spectra onto a logarithmically spaced wavelength grid with a velocity step of 1.5~\kms\ and injecting wavelength-independent random noise to mimic the finite signal-to-noise ratio of real observations. LSD profiles were then computed from these simulated spectra, preserving the 1.5~\kms\ velocity sampling. The noise amplitude was adjusted to achieve a polarimetric precision of $\sigma(V)/I_{\rm c}=10^{-4}$ per 1.5~\kms\ velocity bin in LSD profiles, representative of the higher-quality SPIRou observations of M dwarfs \citep{bellotti:2024,donati:2025a}. With our calculation setup, this corresponds to S/N\,$\approx$\,360 in the input spectra.

Second, the same LSD procedure was applied directly to the full set of local Stokes parameter spectra tabulated as described in Sect.~\ref{sec:prt}. In this case, we used the same line list and telluric exclusion intervals, but retained the velocity sampling of 0.5~\kms\ corresponding to the original resolution of the synthetic spectra. As a final step of the LSD processing of the local Stokes spectra, the mean continuum intensity as a function of limb angle was computed by averaging the Stokes $I$ continuum spectra over wavelength, with the contribution of each spectral line position weighted by the corresponding line depth. The resulting library of local synthetic LSD profiles and average continuum intensities is subsequently used for comparisons with the traditional analytical approximation (Sect.~\ref{sec:UR}), as a basis for computing Stokes profiles that explicitly account for small-scale magnetic fields (Sect.~\ref{sec:PRT}), and for a series of ZDI tests (Sects.~\ref{sec:zdi}, \ref{sec:zdi-weak}, and \ref{sec:zdi-strong}).

The right panels of Fig.~\ref{fig:LSD-mol} present an example of local synthetic LSD profiles in the Stokes $I$ and $V$ parameters, corresponding to the centre of the stellar disk and a line-of-sight magnetic field strength of 100~G. The influence of molecular absorption on the Stokes $I$ LSD profile is clearly visible, whereas its effect on the Stokes $V$ profile is negligible in this weak-field case. Nevertheless, any quantitative interpretation of circular polarisation spectrum, for example a \bz\ measurement, relies on accurate knowledge of the underlying intensity profile. Consequently, the continuum offset introduced by molecular absorption renders such measurements highly uncertain and makes them strongly dependent on the adopted re-normalisation of LSD profiles.

\subsection{Unno-Rachkovsky approximation of local LSD profiles}
\label{sec:UR}

Starting with the work of \citet{morin:2008}, the Unno-Rachkovsky (UR) analytical solution of the polarised radiative transfer equation has been widely employed to interpret LSD Stokes $V$ spectra of M dwarfs. In a recent study, \citet{donati:2025a} extended this approach to the modelling of Stokes $Q$ and $U$ LSD profiles. The UR solution can be derived \citep[e.g.][]{polarization:2004} under the assumptions of a linear variation of the source function with optical depth, a constant absorption matrix, and the absence of any depth dependence of the magnetic field vector or its orientation. In the specific context of stellar LSD profile modelling, an additional assumption is commonly made: the LSD profile is assumed to respond to the magnetic field as a Zeeman triplet with a central wavelength and effective Land\'e factor equal to the average parameters of the LSD line mask ($\lambda_0$ and $z_0$ discussed in Sect.~\ref{sec:lsd}).

Under these assumptions, the normalised Stokes $IQUV$ profiles can be expressed as
\begin{equation}
X (v) / I_{\rm c} = F(\vec{B}, \mu, b, \kappa, v_{\rm D}, v_{\rm L}, v),
\end{equation}
where $X=I$, $Q$, $U$, or $V$, $v$ is the velocity offset from the line centre, $\vec{B}$ is the local magnetic field vector expressed in the observer’s reference frame, $\mu$ is the cosine of the limb angle, $b$ is the slope of the source function with optical depth, $\kappa$ is the line-strength parameter, and $v_{\rm D}$ and $v_{\rm L}$ are the Doppler and Lorentzian widths of the line absorption profile, respectively. In practice, the parameter $b$ is typically fixed to an arbitrary value between 1 and 10, while $\kappa$, $v_{\rm D}$, and $v_{\rm L}$ are determined empirically by fitting the observed LSD profiles.

In practical applications of the UR formalism to stellar ZDI problems, the centre-to-limb variation of the continuum intensity implied by this framework,
\begin{equation}
I_{\rm c} (\mu) / I_{\rm c} (1) = (1+b \mu) / (1+b),
\label{eq:b}
\end{equation}
is usually replaced by a standard linear limb-darkening law,
\begin{equation}
I_{\rm c} (\mu) / I_{\rm c} (1) = 1 - \eta + \eta \mu,
\label{eq:mu}
\end{equation}
where the coefficient $\eta$ is obtained from detailed model atmosphere calculations \citep[e.g.][]{claret:2011}. For the {\sc MARCS} model atmosphere adopted in this work, $\eta=0.21$ at $\lambda=1630.6$~nm.

Departing from this treatment, \citet{bellotti:2024}, following \citet{erba:2024}, proposed fixing the parameter $b$ in the UR formalism according to the relationship between $b$ and $\eta$ implied by Eqs.~(\ref{eq:b}) and (\ref{eq:mu}),
\begin{equation}
b = \eta/(1+\eta).
\end{equation}
This prescription yields significantly smaller values of $b$ than those typically adopted, for example $b=0.27$ for $\eta=0.21$ in the present case. However, this approach is conceptually flawed. Within the UR analytical profile model, $b$ is intended to represent the local slope of the source function $P$, $b=\partial P/\partial\tau$, and its value is therefore expected to differ between the deeper layers where the continuum forms and the higher layers in which spectral lines originate. For the adopted {\sc MARCS} model atmosphere, and assuming the source function to be given by the Planck function at $\lambda=1603.6$~nm, $\partial P/\partial\tau$ increases rapidly from approximately 0.2 at $\log \tau_{\lambda} = 0$ to about 8 at $\log \tau_{\lambda} \approx -2$, where $T \approx T_{\rm eff}$. Therefore, a value of $b$ appropriate for a linear limb-darkening approximation in the continuum is generally unsuitable for describing the centre-to-limb behaviour of spectral lines. In Sect.~\ref{sec:local}, we demonstrate that adopting a small $b$ consistent with $\eta$ in fact exacerbates the discrepancy between the UR approximation and the actual local Stokes LSD profiles.

Finally, ostensibly to account for the presence of unresolved small-scale magnetic fields, \citet{morin:2008} introduced an additional parameter, the global field filling factor $f_{\rm V}$, such that
\begin{equation}
X / I_{\rm c} = f_{\rm V} F(\vec{B}/f_{\rm V}, \mu, b, \kappa, v_{\rm D}, v_{\rm L}).
\end{equation}
With $f_{\rm V} < 1$, this modification implies that the Zeeman splitting and other quantities in the UR model are computed for an amplified magnetic field $\vec{B}/f_{\rm V}$ and the resulting profiles are subsequently rescaled by multiplying by $f_{\rm V}$. For typical values of $f_{\rm V}\approx 0.1$ adopted in M-dwarf ZDI studies, $\vec{B}/f_{\rm V}\gg\vec{B}$. In other words, the line-profile calculations underlying ZDI modelling effectively involve magnetic fields that are much stronger than those shown in the published global field maps.

The physical interpretation of this global field filling factor is unclear. Taken at face value, it can be visualised as a collection of high-contrast, unipolar magnetic spots distributed across the stellar surface in such a way as to reproduce the inferred global field geometry, while occupying only a fraction $f_{\rm V}$ of the surface area \citep{kochukhov:2021,zhang:2025}.

\subsection{Numerical local LSD profiles with small-scale fields}
\label{sec:PRT}

In Sect.~\ref{sec:lsd} we described a procedure for computing local Stokes parameter LSD profiles based on realistic polarised spectrum synthesis over the full studied wavelength range. These profiles, tabulated as $T_X(\vec{B}, \mu, v)$, can in principle be used directly in ZDI inversions, in a manner analogous to magnetic mapping studies of hot stars \citep{kochukhov:2014} or full Stokes vector ZDI of II~Peg by \citet{rosen:2015}. However, applying this approach to the observational modelling of multi-scale magnetic fields in M dwarfs requires the stellar surface to be discretised into an exceptionally large number of elements, as discussed in Sect.~\ref{sec:prt} and illustrated by Fig.~\ref{fig:msup}a. Each surface element must sample a superposition of a smooth global magnetic field, adjusted iteratively by the ZDI inversion, and a small-scale, spatially intermittent field component that is prescribed by Zeeman broadening analysis and held fixed during the global field reconstruction.

While conceptually straightforward, the use of such large surface grids for ZDI inversions is computationally prohibitive in practice, particularly when these inversions must incorporate a large number of rotational phases. To circumvent this limitation, we instead devised a procedure that modifies the local Stokes parameter profiles themselves in order to account for the presence of an additive, unresolved, random small-scale magnetic field component. Specifically, for a given magnetic field strength, orientation, and limb angle, we added a field contribution, $B_{\rm s}\vec{r}$, to a given field vector, $\vec{B}$, and then averaged the resulting Stokes profiles over all possible orientations of the unit vector, $\vec{r}$, uniformly distributed over the full solid angle. This procedure can be expressed as
\begin{equation}
\overline{T_X} (\vec{B}, B_{\rm s}, \mu, v) = \dfrac{1}{4\pi}\int_S T_X (\vec{B} + B_{\rm s} \vec{r},\mu,v) d\Omega,
\label{eq:integ}
\end{equation}
where $d\Omega$ denotes the differential solid angle element. 

This angular averaging effectively captures the contribution of small-scale, unresolved magnetic fields of strength $B_{\rm s}$ to the Zeeman splitting of spectral lines, while producing no net contribution to the polarisation signals themselves. As a result, it closely reproduces the effect of a superposition of a coherent large-scale magnetic field and an isotropic random field component. The principal advantage of applying the transformation defined by Eq.~(\ref{eq:integ}) is that the modified local Stokes profile tables, $\overline{T_X} (\vec{B}, B_{\rm s}, \mu, v)$, can be precomputed prior to the ZDI inversion. These profiles can then be used with standard surface discretisations comprising only a few thousand elements, thereby preserving computational efficiency.

Within this framework, the small-scale field strength, $B_{\rm s}$, can either be treated as a free parameter and adjusted to reproduce the shapes of the observed polarised LSD profiles or fixed using independent constraints from Zeeman broadening analyses. In the latter case, when observational or theoretical considerations indicate the presence of multiple small-scale field components with different characteristic strengths, the formalism can be straightforwardly generalised to account for a multi-component unresolved field by forming a weighted superposition of the corresponding locally averaged profiles,
\begin{equation}
\langle \overline{T_X} (\vec{B}, \mu, v) \rangle = \sum_i f_i \overline{T_X} (\vec{B}, B_{\rm s}^i, \mu, v).
\end{equation}

In the present work, we implemented the numerical evaluation of Eq.~(\ref{eq:integ}) using the quadrature-based spherical integration scheme of \citet{lebedev:1999}. Specifically, we adopted a 5810-point angular grid provided by publicly available software\footnote{\url{https://server.ccl.net/cca/software/SOURCES/FORTRAN/Lebedev-Laikov-Grids}}, which ensures accurate and efficient spherical averaging.

\subsection{ZDI inversions}
\label{sec:zdi}

The ZDI calculations in this study were performed using the inversion code introduced by \citet{kochukhov:2014}. Over the past decade, this software has been extensively applied to magnetic mapping of a wide range of cool stars \citep[e.g.][]{kochukhov:2015a,rosen:2015,lehtinen:2022,hackman:2024,metcalfe:2024} as well as hot stars \citep[e.g.][]{kochukhov:2019,kochukhov:2022,kochukhov:2023a,kochukhov:2025}. A new generation of this inversion code, {\sc InversLSD2}, has recently been developed. The software has been entirely rewritten in modern Fortran90 and includes a number of additional capabilities, such as the modelling of non-radial pulsations and non-spherical stellar geometry associated with rapid rotation. These features, however, are not relevant for the present study.

The inversion code reconstructs the vector magnetic field simultaneously with one scalar parameter at a time, using all four Stokes parameters or selected subsets thereof, based on precomputed tables of local line profiles. This design provides considerable flexibility, allowing the use of different local profile models tailored to specific astrophysical applications. In this work we compared inversions that employ analytical UR local profiles (Sect.~\ref{sec:UR}, hereafter LSD-UR) with calculations based on local LSD profiles derived from detailed polarised radiative transfer and averaged over random small-scale magnetic fields (Sect.~\ref{sec:PRT}, hereafter LSD-PRT). In both approaches, the scalar parameter representing unresolved magnetic structure -- $f_{\rm V}$ in the former case and $B_{\rm s}$ in the latter -- is assumed to be constant across the stellar surface.

In {\sc InversLSD2}, the large-scale magnetic field is described using three families of spherical harmonic functions \citep[see][]{donati:2006b,kochukhov:2014}, yielding a general solenoidal field that includes both poloidal and toroidal components. The inversion proceeds by adjusting the corresponding spherical harmonic expansion coefficients, which are defined for angular degrees $\ell$ ranging from 1 to $\ell_{\rm max}$ and for all azimuthal orders $m=-\ell,-\ell+1\ldots,\ell-1,\ell$ at each degree. For the numerical experiments presented here, we adopted the modified spherical harmonic parameterisation proposed by \citet{lehmann:2022}. This formulation introduces an explicit coupling between the radial and horizontal components of the poloidal field, thereby facilitating the reconstruction of low-order harmonic magnetic geometries.

Because the vast majority of ZDI studies of M dwarfs have relied exclusively on Stokes $V$ LSD profiles, we likewise restricted our magnetic inversions to circular polarisation spectra. The inversions were regularised using a harmonic penalty function \citep{kochukhov:2014}, with the regularisation parameter determined via the step-wise regularisation adjustment procedure discussed by \citet{kochukhov:2017}.

\section{Results}
\label{sec:results}

\subsection{Local LSD Stokes profiles}
\label{sec:local}

\begin{figure*}[t!]
\centering
\includegraphics[width=0.92\hsize]{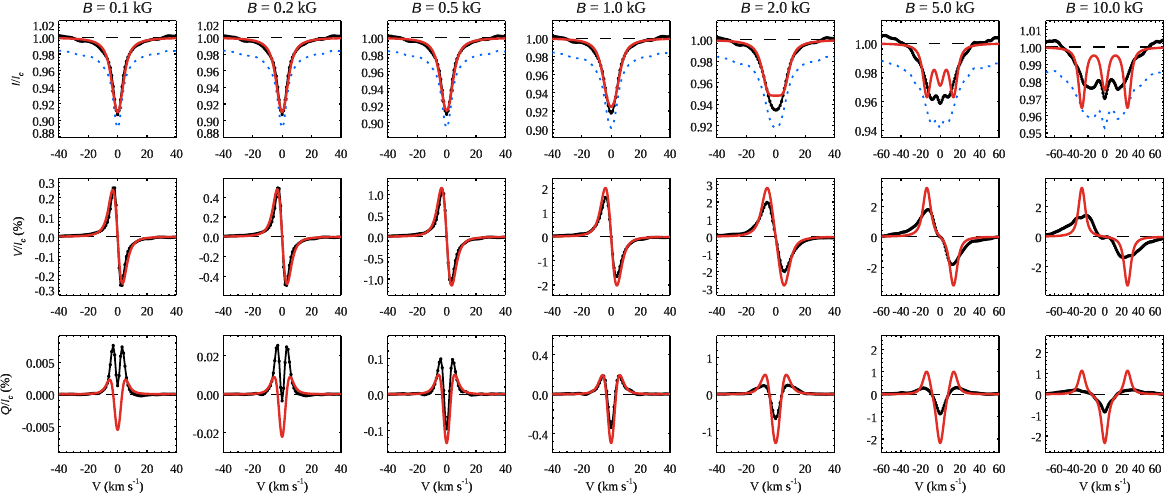}
\caption{Comparison of disk-centre Stokes parameter LSD profiles computed using detailed radiative transfer calculations (symbols connected by solid black lines) and the single-line UR approximation (red lines). The dotted blue lines show LSD Stokes $I$ profiles before continuum re-normalisation. Each row displays a different Stokes parameter: $I$ (\textit{top}), $V$ (\textit{middle}), and $Q$ (\textit{bottom}). The columns correspond to increasing magnetic field strengths, ranging from 100 G to 10 kG. In all panels, the magnetic field vector is inclined at an angle of 45\degr\ to the line of sight.}
\label{fig:LSD-UR}
\end{figure*}

\begin{figure*}[t!]
\centering
\includegraphics[width=0.92\hsize]{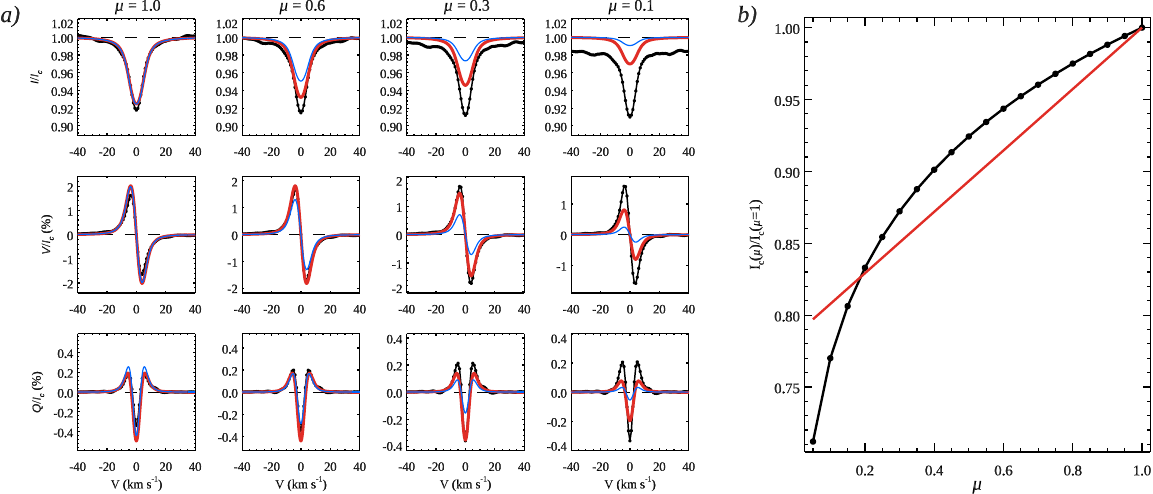}
\caption{Centre-to-limb variation of LSD line profiles (\textit{a}) and continuum intensity (\textit{b}). Panel (a) compares Stokes $IVQ$ LSD profiles computed with full radiative transfer calculations (black symbols connected with solid lines) to those obtained using the single-line UR approximation for $b = 5$ (thick red lines) and $b = 0.27$ (thin blue lines). The latter value is consistent with the continuum limb-darkening coefficient $\eta = 0.21$. In all cases the magnetic field strength is 1 kG and the field vector is inclined by 45\degr\ relative to the line of sight. Panel (b) compares the continuum limb-darkening predicted by detailed radiative-transfer calculations (black symbols connected with solid lines) with a linear limb-darkening law using $\eta = 0.21$ (red line).}
\label{fig:LSD-UR-mu}
\end{figure*}

\begin{figure*}[t!]
\centering
\includegraphics[height=6.76cm]{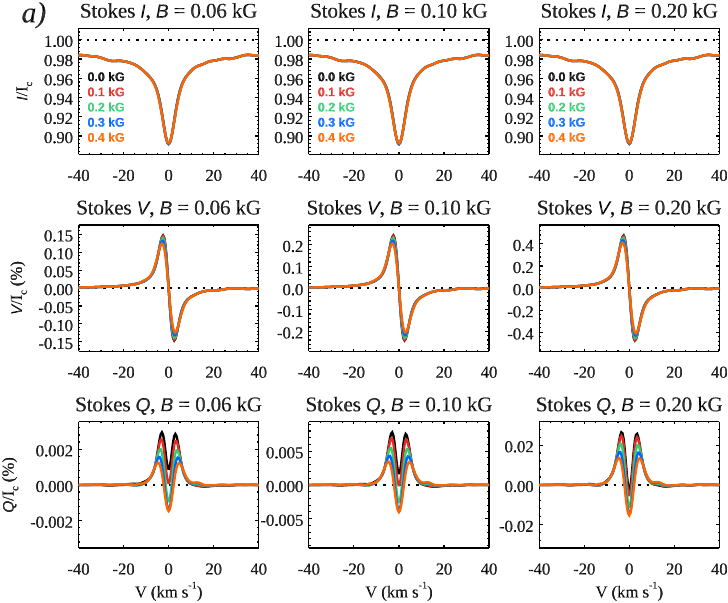}\hspace*{7mm}
\includegraphics[height=6.76cm]{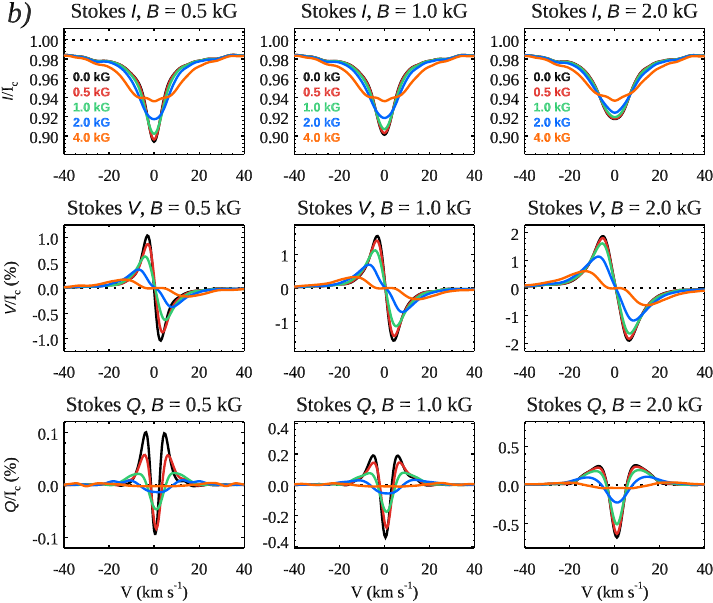}
\caption{Effect of isotropic small-scale magnetic fields with different strengths on the disk-centre Stokes $IVQ$ LSD profiles. Each column presents the Stokes parameter profiles for a given local field strength, $B$, and a 45\degr\ inclination relative to the line of sight. Different colours correspond to increasing strength of the small-scale field, $B_{\rm s}$. \textit{Panel (a)}: LSD Stokes profiles for inactive stars ($B=60$--200~G, $B_{\rm s}=100$--400~G). \textit{Panel (b)}: LSD Stokes profiles for active stars ($B=0.5$--2.0~kG, $B_{\rm s}=0.5$--4.0~kG).}
\label{fig:LSD-Bran}
\end{figure*}

We begin by comparing local Stokes parameter profiles computed using detailed spectrum synthesis (LSD-PRT) with those obtained using the UR approximation (LSD-UR). Figure~\ref{fig:LSD-UR} presents a set of Stokes $I$, $V$, and $Q$ profiles for magnetic field strengths ranging from 0.1 to 10~kG, assuming a field vector inclination of 45\degr\ with respect to the line of sight and a disk-centre position on the stellar surface ($\mu=1$). For this magnetic field geometry, the signal in Stokes $U$ is very weak. Consequently, Stokes $U$ profiles are not shown in Fig.~\ref{fig:LSD-UR}.

As discussed earlier, the presence of numerous molecular blends depresses the continuum level around the LSD profile calculated with an atomic line mask. This effect necessitates re-normalisation of the profiles before applying the UR formalism. In this case, all profiles were multiplied by a factor of 1.018 to restore the Stokes $I$ wings to unity at velocity offsets of $\pm$\,40~\kms. Following this re-normalisation, the line strength and broadening parameters were adjusted by fitting the non-magnetic Stokes $I$ profile. This procedure yielded $\kappa=0.83$, $v_{\rm D}=1.08$~\kms, and $v_{\rm L}=3.35$~\kms. These parameters were then held fixed for the computation of profiles with $B\ne0$.

The resulting comparison between the LSD-PRT and LSD-UR profiles shows that the analytical approximation is successful under certain conditions, while exhibiting significant limitations outside those regimes. For magnetic field strengths below 1~kG, the UR profiles closely reproduce the LSD-PRT results for both Stokes $I$ and $V$. However, for stronger fields, systematic differences in the shapes and amplitudes of the Stokes $I$ and $V$ profiles become apparent. In particular, the UR approximation tends to underestimate the profile widths while overestimating the peak amplitudes of the polarisation signals. The severity of these discrepancies increases progressively with magnetic field strength: while the disagreement remains moderate at $B=2$~kG, the two sets of profiles differ substantially at $B=10$~kG.

The performance of the UR approximation is notably poorer for the linear polarisation LSD spectra. This behaviour is not unexpected, since the profile self-similarity that underpins the LSD-UR treatment of Stokes $V$ in the weak-field regime does not apply to Stokes $Q$ and $U$ to the same extent \citep{polarization:2004,kochukhov:2010a}. Our analysis indicates that the analytical Stokes $Q$ profiles shown in Fig.~\ref{fig:LSD-UR} provide a reasonable approximation to the LSD-PRT calculations only in a narrow range around $B\approx1$~kG. For both weaker and stronger magnetic fields, the disagreement increases significantly.

We next examine the centre-to-limb behaviour of the LSD Stokes profiles and the continuum intensity, as shown in Fig.~\ref{fig:LSD-UR-mu}. For these calculations, the magnetic field strength was fixed at 1~kG, while the cosine of the limb angle, $\mu$, was varied from 1 to 0.1. The same line parameters and profile re-normalisation as in the previous experiment were adopted. Two sets of UR profiles were considered: one computed using an arbitrary value of $b=5$, and another using $b=0.27$, inferred from the linear limb-darkening coefficient $\eta=0.21$ as discussed in Sect.~\ref{sec:UR}. The corresponding linear limb-darkening curve is compared with that obtained from detailed radiative transfer calculations in Fig.~\ref{fig:LSD-UR-mu}b, which reveals substantial differences between the two approaches.

The LSD-PRT profiles shown in Fig.~\ref{fig:LSD-UR-mu}a exhibit nearly constant strengths in both intensity and polarisation as a function of $\mu$. In contrast, the UR approximation with $b=5$ predicts progressively weaker line profiles towards the stellar limb. Adopting $b=0.27$ further amplifies this trend, thereby increasing the discrepancy between the LSD-PRT and LSD-UR results. An additional effect visible in Fig.~\ref{fig:LSD-UR-mu}a, which cannot be captured by analytical methods, is the increasing continuum offset at lower $\mu$. As the line of sight approaches the stellar limb, line formation shifts to cooler atmospheric layers, leading to enhanced molecular absorption and, consequently, a larger continuum offset in the Stokes $I$ profiles.

Finally, we assessed the impact of the transformation applied to the Stokes parameter profiles, as described in Sect.~\ref{sec:UR}. This procedure modifies the shapes of the LSD-PRT profiles in order to account for an isotropic small-scale background magnetic field component. Two regimes are considered for the profiles shown in Fig.~\ref{fig:LSD-Bran}: a weak-field case with $B=60$--200~G and $B_{\rm s}=100$--400~G, and a strong-field scenario with $B=0.5$--2.0~kG and $B_{\rm s}=0.5$--4.0~kG. The former corresponds to a typical inactive M dwarf, while the latter is representative of an active star.

Figure~\ref{fig:LSD-Bran}a demonstrates that the inclusion of realistic levels of small-scale magnetic field produces only minor changes in the Stokes $I$ and $V$ profiles in the weak-field regime. In contrast, the Stokes $Q$ profiles are systematically modified, following the evolution observed for magnetic field strengths between $B=100$~G and 500~G in Fig.~\ref{fig:LSD-UR}. In the strong-field case illustrated in Fig.~\ref{fig:LSD-Bran}b, the transformation with $B_{\rm s}>B$ leads to pronounced changes in both intensity and polarisation profiles. As expected, the Stokes $I$ profiles broaden as a result of enhanced Zeeman splitting. The Stokes $V$ and $Q$ profiles also become broader and decrease in amplitude, with the linear polarisation signals being affected more strongly. Nevertheless, despite these substantial modifications, the Stokes $V$ profiles retain the same net circular polarisation signature, corresponding to the longitudinal magnetic field, as in the original unmodified LSD-PRT spectra.

\subsection{ZDI of an inactive M dwarf}
\label{sec:zdi-weak}

\begin{figure*}[t!]
\centering
\includegraphics[height=8.5cm]{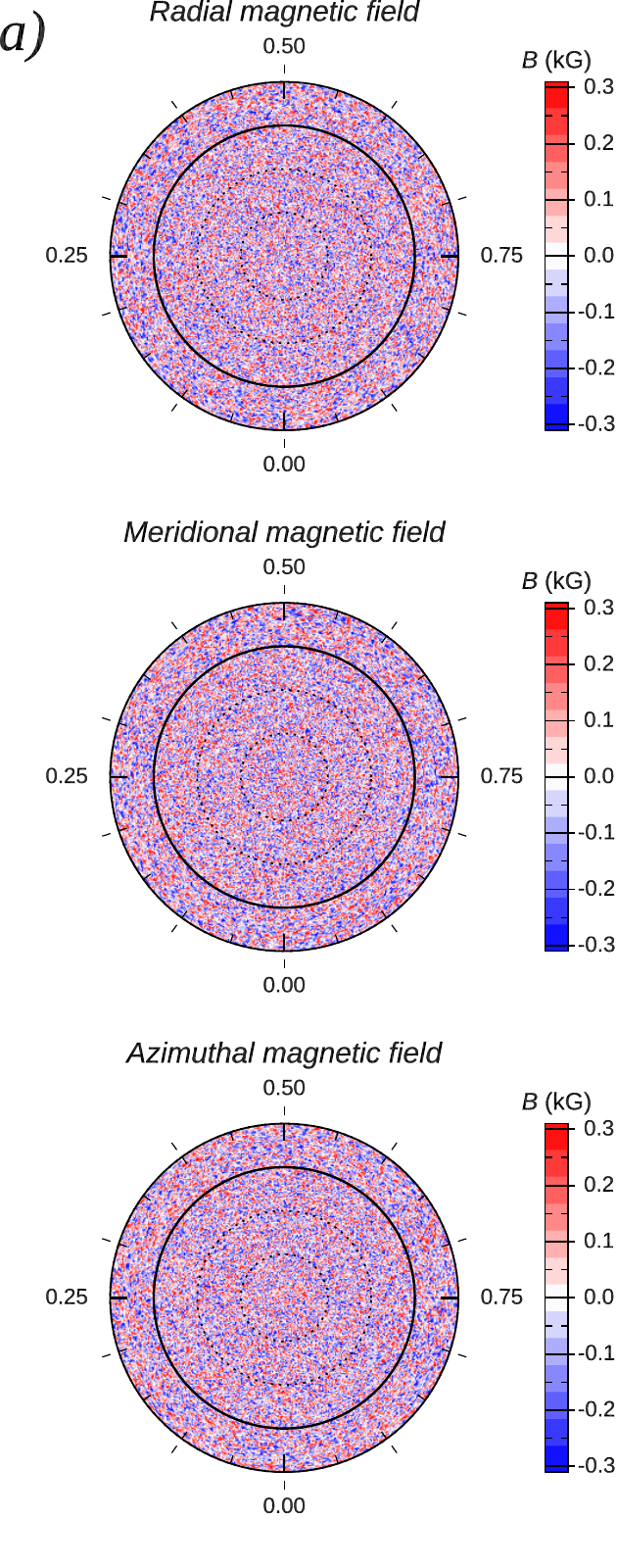}\hspace*{2mm}
\includegraphics[height=8.5cm]{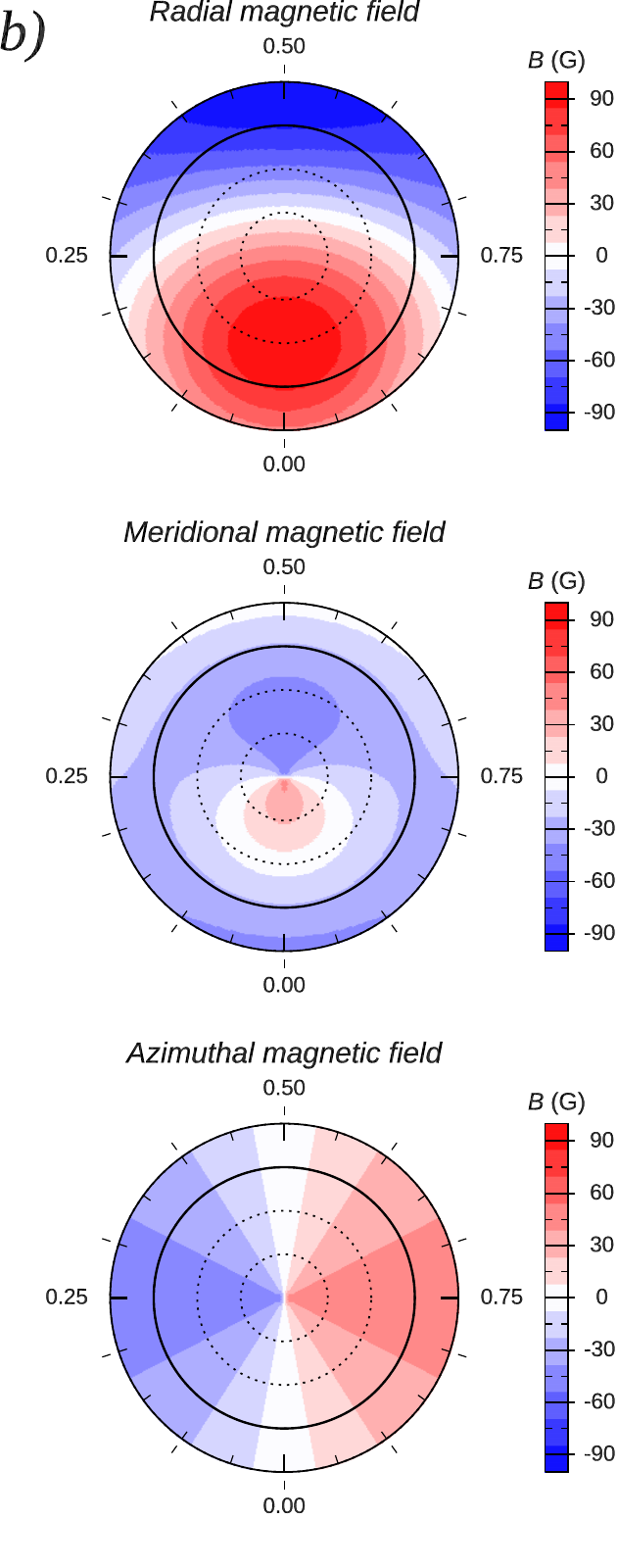}\hspace*{2mm}
\includegraphics[height=8.5cm]{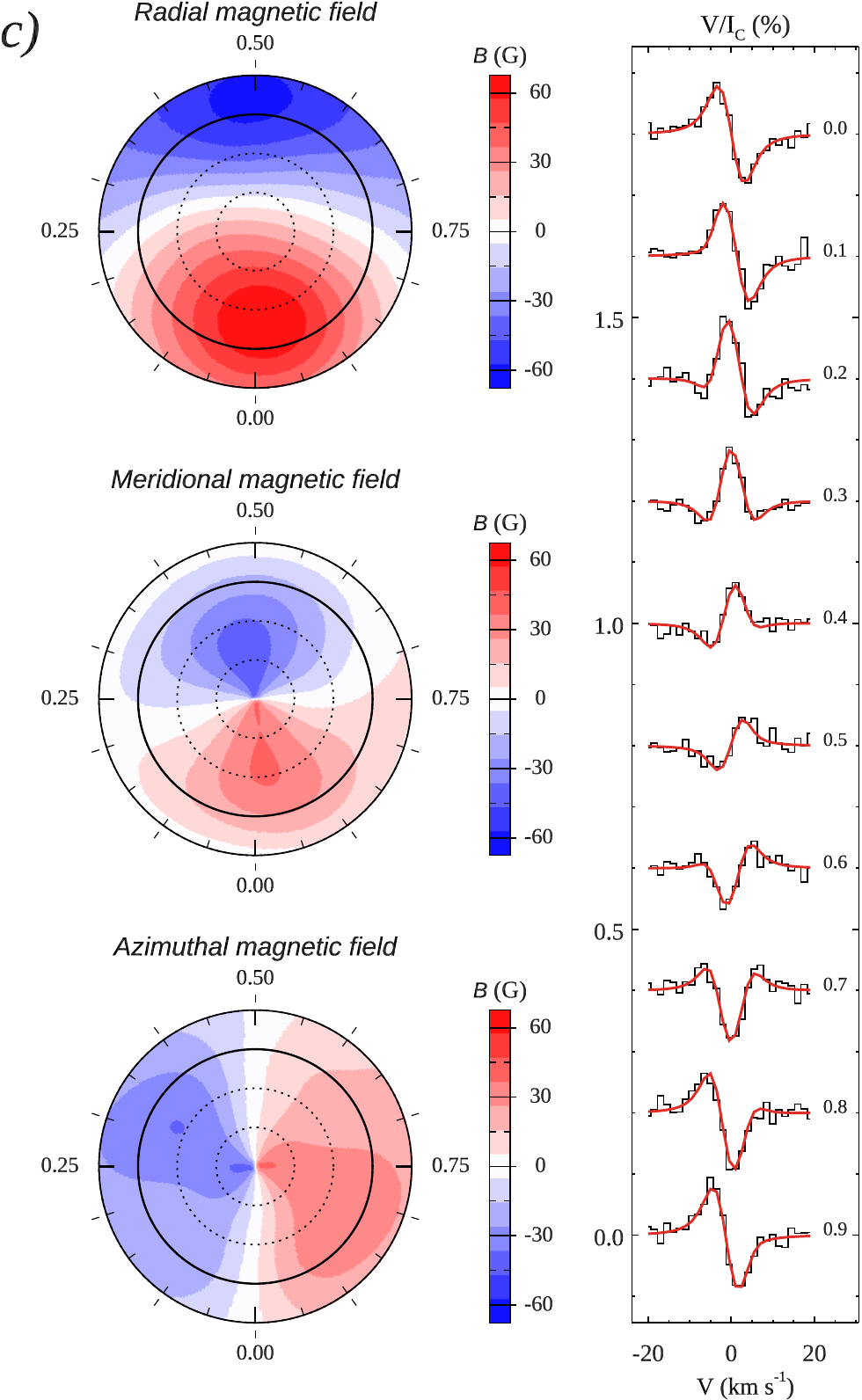}\hspace*{2mm}
\includegraphics[height=8.5cm]{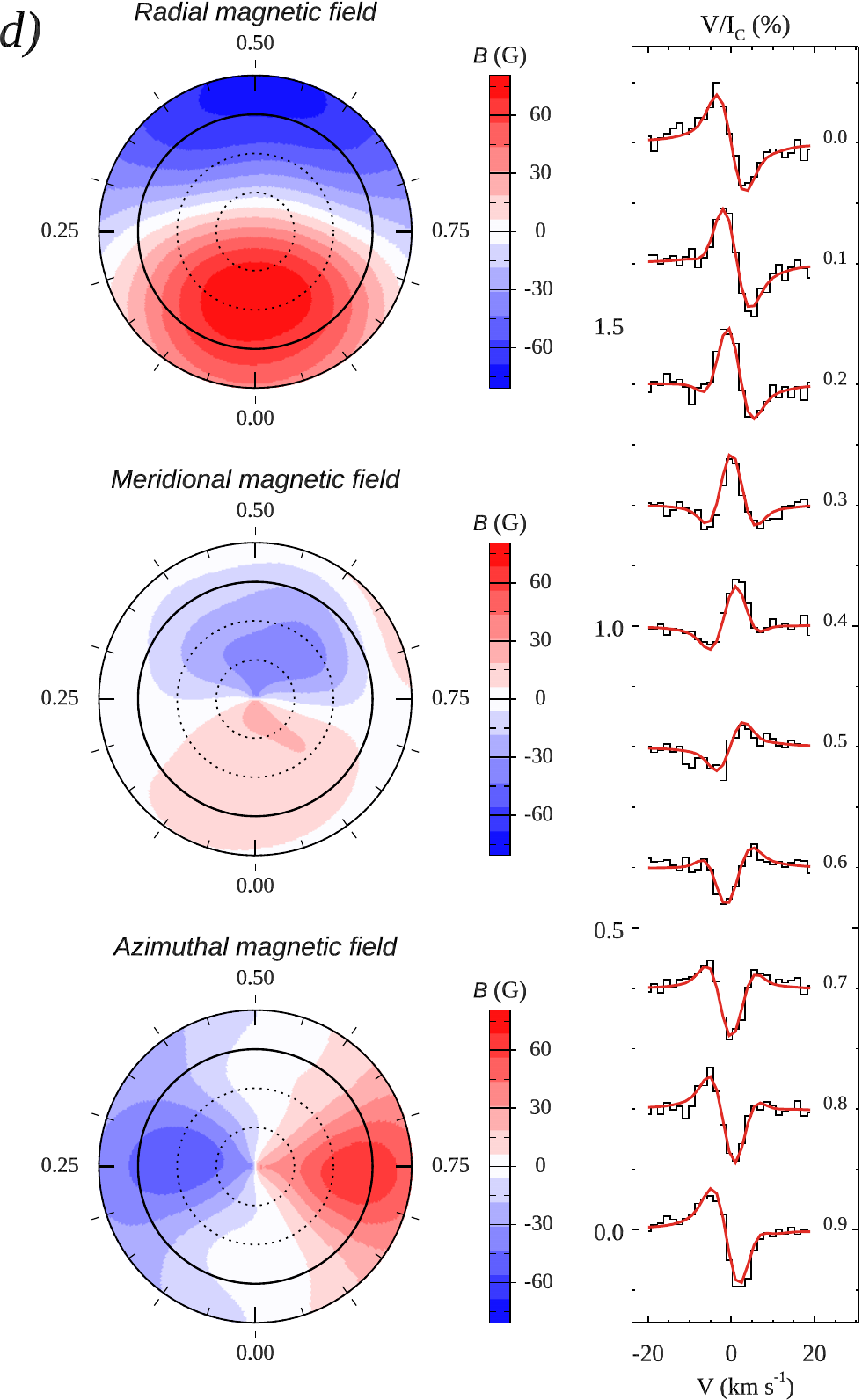}
\caption{Comparison between the input small-scale (\textit{a}) and global (\textit{b}) magnetic field geometry and ZDI reconstructions using two approaches: the single-line LSD-UR profile approximation (\textit{c}) and a multi-line numerical model incorporating a small-scale field (\textit{d}; LSD-PRT). Each panel presents flattened polar projections of the radial, meridional, and azimuthal components of the small-scale (a) or global (b-d) magnetic field. 
The thick circle marks the stellar equator, with dotted lines drawn every 30\degr\ in latitude. 
The right columns of panels (c) and (d) show the simulated observed Stokes $V$ spectra (black histograms) alongside the corresponding ZDI fits (solid red lines). 
Profiles are vertically offset according to rotational phase, with the corresponding phases listed to the right of the profiles.
For this test, the star was modelled with a global dipolar field of strength $B_{\rm d} = 100$ G and obliquity $\beta = 60\degr$, along with an isotropic small-scale magnetic component of $B_{\rm s} = 300$ G.
}
\label{fig:ZDI1}
\end{figure*}

We conclude the numerical investigations presented in this paper by simulating ZDI reconstruction of a global magnetic field in the presence of a dominant small-scale magnetic component. In the first experiment, we adopted magnetic field parameters representative of a weakly active M dwarf \citep{reiners:2022,lehmann:2024}. The small-scale random field was set to $B_{\rm s}=300$~G, while the large-scale field was represented by a dipole of strength $B_{\rm d}=100$~G, inclined by $\beta=60\degr$ with respect to the stellar rotation axis. These two magnetic components are illustrated in Fig.~\ref{fig:ZDI1}a and b. Using this composite magnetic configuration, we simulated observed LSD profiles following the procedure described in Sect.~\ref{sec:lsd}, assuming an inclination of $i=60\degr$, a projected rotational velocity of \vsini\,=\,5~\kms, and a rotational phase coverage consisting of ten observations. We then performed ZDI reconstructions using both the LSD-UR and LSD-PRT approaches. Although the input large-scale field in this test was purely dipolar, the ZDI inversions were carried out with {\sc InversLSD2} allowing for spherical harmonics up to $\ell_{\rm max}=5$ and treating all three sets of harmonic coefficients as free parameters. To disentangle systematic biases from the effects of noise in the input data, each inversion was repeated 50 times using different random noise realisations.

\begin{table*}
\caption{ZDI results for inactive and active M-dwarf scenarios with different LSD profile modelling approaches. \label{tbl:results}}
\centering
\begin{tabular}{l | c c c | c c c}
\hline\hline
 & \multicolumn{3}{c|}{Inactive M dwarf} & \multicolumn{3}{c}{Active M dwarf} \\
 & Input & \multicolumn{2}{c|}{ZDI reconstruction} & Input & \multicolumn{2}{c}{ZDI reconstruction}  \\
Parameter &  & LSD-UR & LSD-NRT &  & LSD-UR & LSD-NRT \\
\hline
$\langle B_{\rm V} \rangle$ (G) & 69 & $45\pm2$ & $55\pm2$ & 690 & $522\pm26$ & $546\pm24$ \\
$B_{\rm max}$ (G)  & 100 & $73\pm4$ & $87\pm4$ & 1000 & $1177\pm78$ & $874\pm35$ \\
$B_{\rm d}$ (G) & 100 & $62\pm2$ & $79\pm3$ & 1000 & $594\pm71$ & $802\pm32$ \\
$\beta$ (\degr) & 60 & $66\pm3$ & $65\pm3$ & 60 & $88\pm16$ & $65\pm4$ \\
$B_{\rm s}$ (G) & 300 & -- & 300$^\dagger$ & 3000 & -- & $3014\pm14$ \\
$f_{\rm V}$ & -- & 1.00$^\dagger$ & -- & -- & $0.15\pm0.02$ & -- \\
$E_{\rm pol}$ (\%) & 100 & $91\pm3$ & $95\pm2$ & 100 & $88\pm3$ & $96\pm1$ \\
$E_{m=0}$ (\%) & 25 & $12\pm3$ & $15\pm5$ &  25 & $35\pm6$ & $15\pm6$ \\
$E_{\ell=1}$ (\%) & 100 & $96\pm2$ & $95\pm2$ & 100 & $78\pm4$ & $98\pm1$ \\
RMS($|\vec{B}_{\rm V}-\vec{B}_{\rm input}|$) (G) & -- & $35\pm3$ & $27\pm3$ & -- & $650\pm124$ & $240\pm45$  \\
\hline
\end{tabular}
\tablefoot{$^\dagger$Fixed parameter in the ZDI inversion.}
\end{table*}

\begin{figure*}[t!]
\centering
\includegraphics[height=8.5cm]{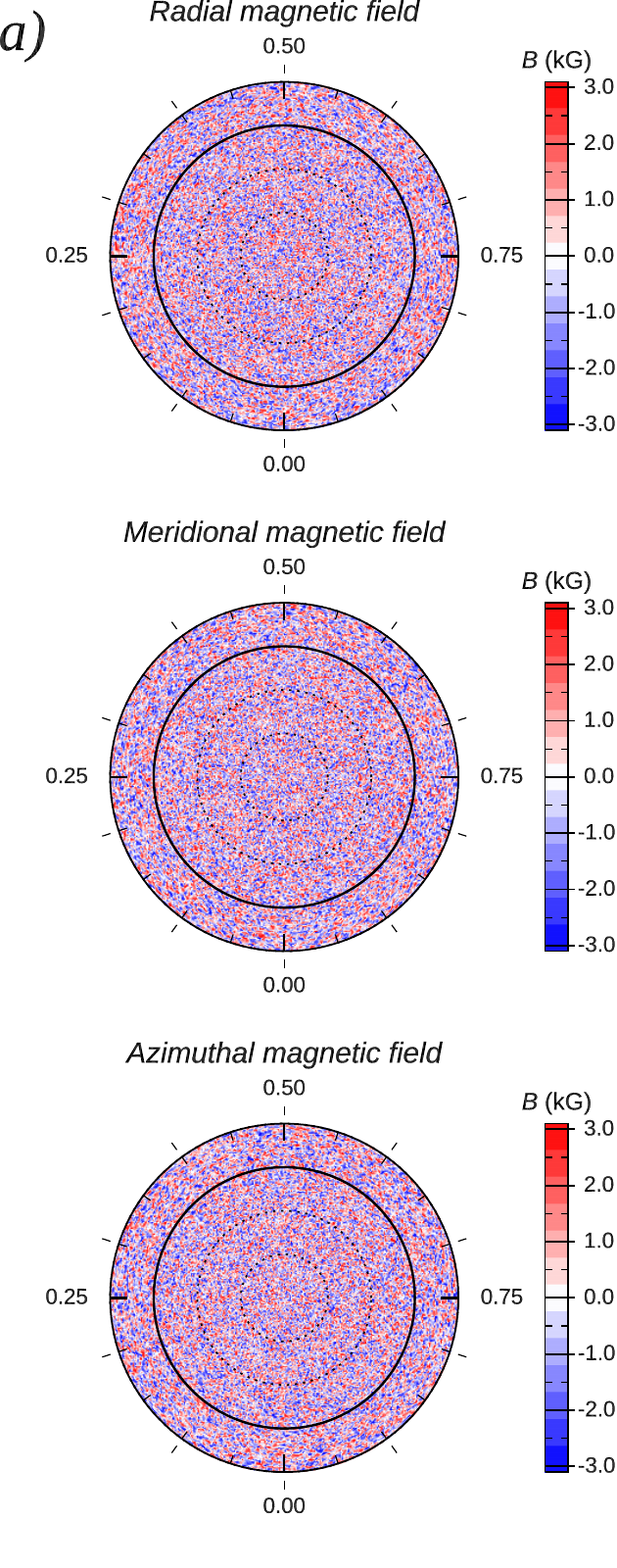}\hspace*{2mm}
\includegraphics[height=8.5cm]{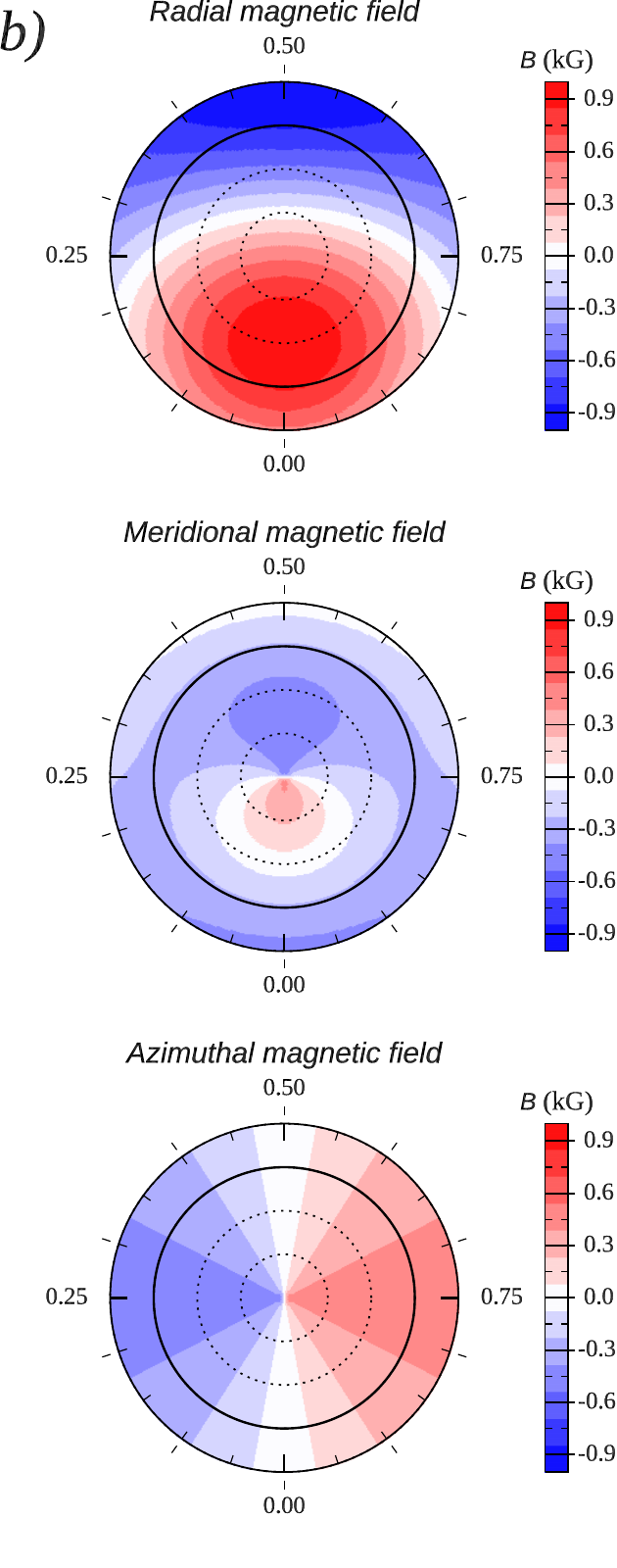}\hspace*{2mm}
\includegraphics[height=8.5cm]{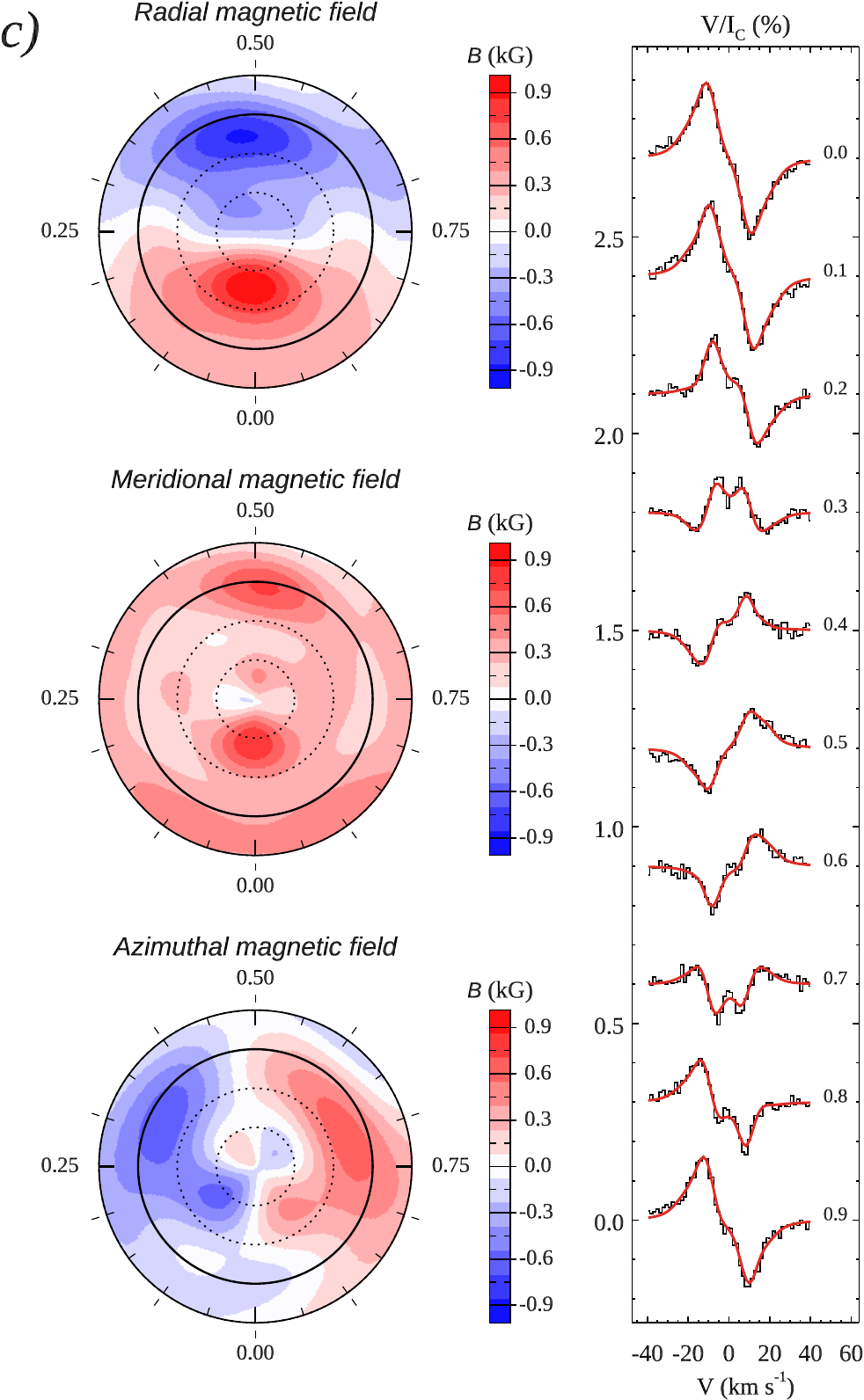}\hspace*{2mm}
\includegraphics[height=8.5cm]{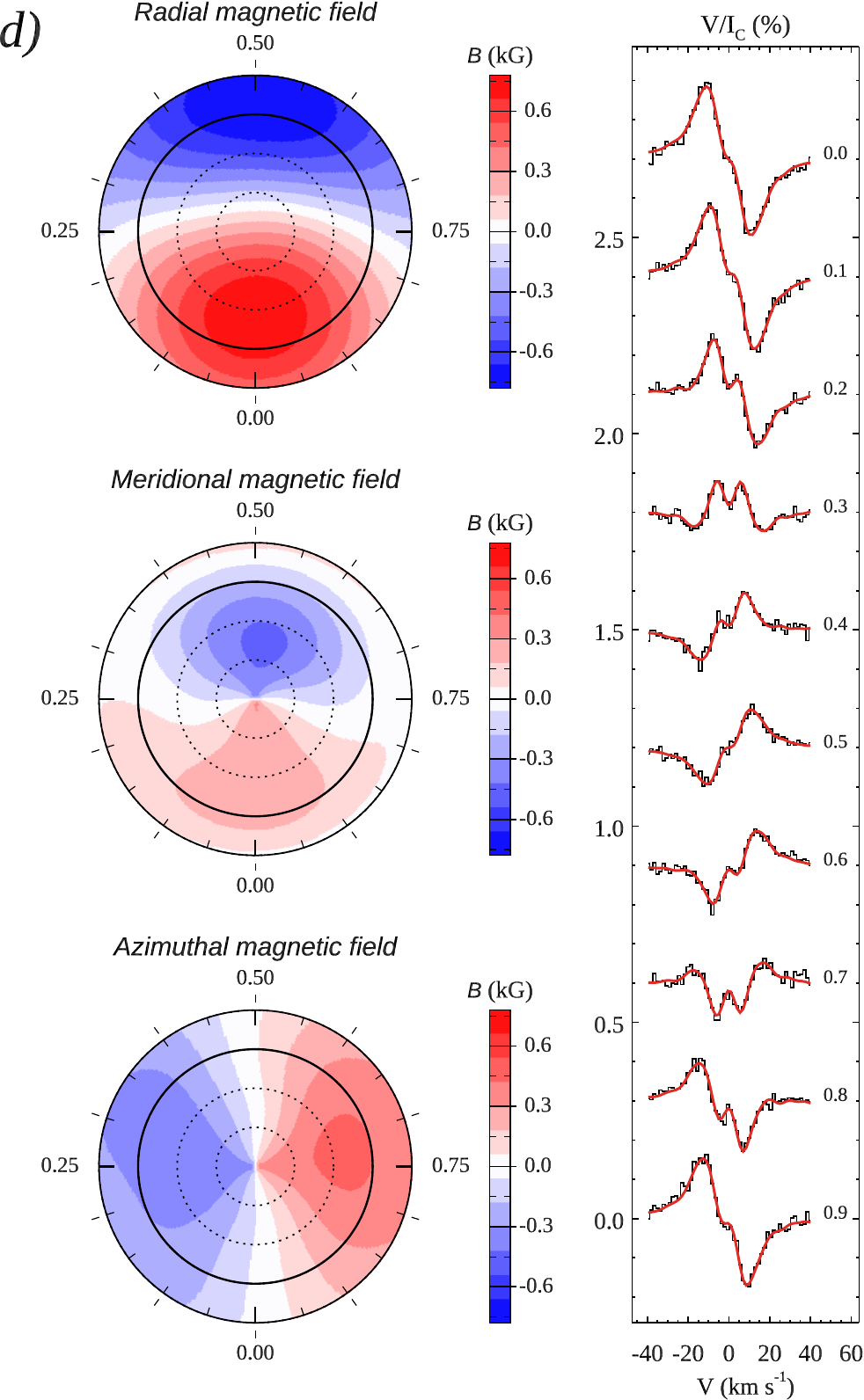}
\caption{Same as Fig.~\ref{fig:ZDI1} but for the star with the input global field $B_{\rm d}=1$~kG and the small-scale field $B_{\rm s}=3$~kG. The best-fitting global field filling factor for the ZDI reconstruction in panel (c) was $f_{\rm V} = 0.15$, while the optimal small-scale field for the reconstruction in panel (d) was 3.0 kG.}
\label{fig:ZDI2}
\end{figure*}

Representative reconstructed global magnetic field maps obtained from single inversions with the two LSD profile modelling methods are shown in Fig.~\ref{fig:ZDI1}c and d, together with the corresponding fits to the simulated observations. A quantitative comparison between the input and reconstructed magnetic field parameters is provided in Table~\ref{tbl:results}. The reported values correspond to the mean and one standard deviation derived from the ensemble of 50 ZDI reconstructions. The table summarises global field properties commonly discussed in observational studies, including the average global field strength ($\langle B_{\rm V} \rangle$), the maximum field strength ($B_{\rm max}$), and the dipolar field strength ($B_{\rm d}$) and obliquity ($\beta$) inferred from the $\ell=1$ poloidal harmonic coefficients. We also list the fractional contributions of the poloidal ($E_{\rm pol}$), axisymmetric ($E_{m=0}$), and dipolar ($E_{\ell=1}$) components to the total magnetic energy. In addition, we computed the root-mean-square (RMS) of the modulus of the vector difference between the true and recovered global fields as an overall metric of the reconstruction quality.

As illustrated in Fig.~\ref{fig:ZDI1} and summarised in Table~\ref{tbl:results}, both the LSD-UR and LSD-PRT methods provide a satisfactory reconstruction of the global magnetic field. In both cases, the recovered field strength is systematically underestimated, although this bias is less pronounced for the LSD-PRT approach. We verified that this underestimation arises solely from the redistribution of magnetic energy in the reconstructed maps, whereby power diffuses from the $\ell=1$ mode to higher-order harmonics and from poloidal to toroidal components. If the inversion is restricted to $\ell=1$ and only poloidal harmonic coefficients are allowed to vary, {\sc InversLSD2} achieves an almost perfect recovery of the input dipolar field with either LSD modelling approach.

In the field strength regime explored in this ZDI experiment, the Stokes $V$ profiles scale linearly with magnetic field strength, consistent with the weak-field approximation, while the Stokes $I$ spectra remain essentially unaffected by the magnetic field. Under these conditions, the parameters of the small-scale magnetic field do not influence the ZDI reconstruction and cannot be constrained by it. Accordingly, we fixed $f_{\rm V}=1$ for the LSD-UR inversions and $B_{\rm s}=300$~G for the LSD-PRT inversions. At the same time, the LSD-UR results remain largely unchanged if one assumes $f_{\rm V}=0.1$--0.15 in the magnetic mapping, because the corresponding maximum field strength, $B_{\rm max}/f_{\rm V}$, stays below the $\approx$\,1~kG threshold at which the UR approximation for local LSD profiles begins to break down.

\subsection{ZDI of an active M dwarf}
\label{sec:zdi-strong}

The second set of ZDI inversions explored a substantially stronger magnetic field regime, representative of an active M dwarf \citep[e.g.][]{morin:2008,shulyak:2017}. In this experiment, we adopted a small-scale magnetic field of strength $B_{\rm s}=3$~kG combined with a global dipolar component characterised by $B_{\rm d}=1$~kG and an obliquity of $\beta=60\degr$. To facilitate a direct comparison with the weak-field tests presented in Sect.~\ref{sec:zdi-weak}, we retained the same stellar inclination, projected rotational velocity \vsini, and rotational phase sampling. In contrast to the previous case, however, the resulting total field strength of up to $\sqrt{B_{\rm s}^2 + B_{\rm d}^2}\approx 3.2$~kG strongly affects the shapes of both Stokes $I$ and $V$ profiles. This necessitates the use of a $f_{\rm V}<1$ filling factor within the LSD-UR framework and the adoption of a non-negligible $B_{\rm s}$ in the LSD-PRT approach. Because these small-scale field parameters would not be known a priori in real ZDI applications, we treated them as free parameters. Each of the 50 ZDI reconstructions was therefore repeated multiple times for different values of $f_{\rm V}$ and $B_{\rm s}$. We explored the range $f_{\rm V}=0.1$--0.3 and varied $B_{\rm s}$ between 2.5 and 3.5~kG, selecting the optimal parameters based on the quality of the fit to the simulated Stokes $V$ observations.

The adopted input geometries of the small- and large-scale magnetic field components are shown in Figs.~\ref{fig:ZDI2}a and b. Representative results of individual LSD-UR and LSD-PRT inversions are displayed in Figs.~\ref{fig:ZDI2}c and d, respectively. A quantitative comparison of the input and reconstructed magnetic field properties is provided in Table~\ref{tbl:results}.

Both the LSD-UR and LSD-PRT methods reproduce the simulated observational data with high fidelity, achieving comparably good fits to the LSD Stokes profiles. However, in contrast to the weak-field experiments discussed in Sect.~\ref{sec:zdi-weak}, the inferred magnetic field topologies and their physical interpretation differ markedly between the two approaches. The LSD-PRT inversion yields a faithful reconstruction of the input large-scale oblique dipolar field and successfully recovers the correct small-scale field strength $B_{\rm s}$. Although it slightly underestimates $\langle B_{\rm V} \rangle$, $B_{\rm max}$, and $B_{\rm d}$, it reproduces the correct dipolar obliquity $\beta$ and correctly identifies the global field as predominantly dipolar and poloidal.

In contrast, the LSD-UR inversion produces a significantly more complex and distorted global field geometry. The reconstructed maps exhibit enhanced spurious $\ell>1$ harmonic modes, a non-negligible toroidal component, and an incorrect sign of the meridional field. These artefacts are consistently present in the majority of test inversions, independent of the noise realisation. While Fig.~\ref{fig:ZDI2}c still suggests that the overall bipolar structure and the order of magnitude of the field strength are broadly recovered, the inferred field inclination is severely overestimated, and the RMS deviation from the true field is a factor of 2.7 larger than in the LSD-PRT reconstruction. 

The LSD-UR ZDI converges on a filling factor of $f_{\rm V}=0.15\pm0.02$, closely resembling values commonly reported in observational studies. However, the physical interpretation implied by this result -- namely, a network of magnetic spots covering approximately 15 per cent of the stellar surface with local field strengths reaching $B_{\rm max}/f_{\rm V}\approx7.8$~kG -- is entirely inconsistent with the actual composite magnetic field topology used to generate the synthetic observations.

Within the framework of the small- and large-scale magnetic field superposition model adopted here, the failure of the LSD-UR ZDI can be traced to a mismatch between the actual LSD profiles and the UR approximation, as well as to the inability of the filling-factor formalism to reproduce the impact of the small-scale field on the Stokes parameter profiles. The severity of these shortcomings depends on the total magnetic field strength, which is primarily determined by the small-scale component. Consequently, we expect traditional LSD-UR inversions to encounter similar difficulties for both weak and strong global fields when the dominant small-scale field has a comparable strength in the two cases. This behaviour was confirmed by a test inversion analogous to those discussed in this section, but employing $B_{\rm d}=200$~G together with $B_{\rm s}=3.0$~kG.

\section{Summary and discussion}
\label{sec:disc}

In this paper we have presented a comprehensive investigation of the use of LSD Stokes parameter profiles for diagnosing stellar magnetic fields. Our study builds on the work of \citet{kochukhov:2010a}, extending it to the context of near-infrared spectropolarimetry of low-mass stars and systematically assessing how different local LSD profile modelling approaches affect the outcome of ZDI inversions. The foundation of our analysis is a set of theoretical Stokes parameter spectra, computed with the highest level of physical realism currently achievable using one-dimensional M dwarf model atmospheres. On the basis of these calculations, we critically examined the widely used interpretation of LSD profiles based on the Unno-Rachkovsky analytical solution of the polarised radiative transfer equation combined with treating LSD profiles as a single Zeeman triplet (the LSD-UR approach). We contrasted this traditional framework with a more advanced treatment of local LSD profiles that relies on grids of synthetic spectra, explicitly accounting for background molecular opacity and incorporating the effects of an isotropic small-scale random magnetic field (the LSD-PRT method). Both approaches were subsequently implemented in ZDI and tested using simulated observations based on realistic superpositions of small-scale and global magnetic field components representative of M dwarfs spanning a range of magnetic activity levels.

Our main results can be summarised as follows.
\begin{enumerate}[noitemsep, topsep=0.5pt]
\item The LSD-UR method provides an adequate description of Stokes $I$ and $V$ LSD profiles for magnetic field strengths up to $\approx$\,1~kG. For stronger fields, this approximation breaks down and is, in general, not applicable to Stokes $Q$ and $U$ LSD profiles.
\item The LSD-UR approach fails to reproduce the centre-to-limb variation of LSD Stokes profiles. Adopting UR analytical model parameters that are consistent with an externally imposed linear limb-darkening coefficient \citep{bellotti:2024} further exacerbates this discrepancy.
\item Despite these limitations, the use of LSD-UR local profile modelling in ZDI is justified for magnetically inactive stars, provided that the total field strength does not exceed a few hundred gauss. In this regime, introducing a global field filling factor ($f_{\rm V}$) is unnecessary.
\item In contrast, applying the LSD-UR method to ZDI of strongly magnetic M dwarfs, particularly when fitting for $f_{\rm V}$, is prone to recovering a distorted large-scale field topology and leads to a spurious interpretation involving unrealistically strong magnetic fields confined to a small fraction of the stellar surface. Our newly developed LSD-PRT local profile modelling approach avoids these shortcomings and yields accurate and robust ZDI results for both weakly and strongly magnetic M dwarfs.
\end{enumerate}

\smallskip

In light of our results, it is necessary to critically re-evaluate ZDI studies of M dwarfs that relied on the LSD-UR line profile treatment in combination with the filling factor formalism. In Appendix~\ref{sec:summary} we have compiled all such studies published between 2008 and 2025, covering both optical and near-infrared analyses. While our calculations do not directly address the optical regime, the wavelength scaling of Zeeman splitting relative to the intrinsic line width, combined with earlier work \citep{kochukhov:2010a}, suggests that the LSD-UR approach in the optical is likely valid only up to fields of $\approx$2~kG.

A key parameter determining the reliability of these inversions is the maximum local field strength, $B_{\rm max}/f_{\rm V}$, employed in ZDI. Although rarely discussed explicitly in the literature, this quantity can be calculated from published $B_{\rm max}$ values or using the maximum field inferred from published ZDI maps. Table~\ref{tbl:zdi} reveals that in many cases, $B_{\rm max}/f_{\rm V}$ reaches 5--10~kG and, in the extreme cases of EV~Lac, YZ~CMi, GJ~51, and WX UMa, 20--40 kG \citep{morin:2008,morin:2010}. The implied picture of small, unipolar magnetic spots occupying 10--15\% of the stellar surface and harbouring such extreme field strengths is physically questionable and has not been definitively supported by direct observational evidence, either through the detection of small, high-contrast magnetic spots in the spectra of rapidly rotating M dwarfs or through the unambiguous identification of $\gg$\,10~kG magnetic field components in Zeeman broadening studies.

Moreover, even in the context of polarisation profile modelling of moderately strong fields, our results demonstrate that the LSD-UR approximation becomes fundamentally invalid once the true local field strength, $B_{\rm max}/f_{\rm V}$, significantly exceeds 1~kG, implying that the majority of the ZDI maps summarised in Table~\ref{tbl:zdi} are subject to significant artefacts and systematic biases. The only exceptions are the analyses of DX~Cnc \citep{morin:2010}, DS~Leo \citep{bellotti:2024}, and several inactive M dwarfs by \citet{lehmann:2024}, for which $B_{\rm max}/f_{\rm V}$ generally remained below 1.5--2~kG.

Consequently, it is difficult -- if not impossible -- to assess which conclusions drawn from previous M-dwarf ZDI studies are trustworthy and which may be compromised by the limitations of the LSD-UR framework. A definitive resolution requires revisiting these inversions using the improved LSD Stokes profile treatment introduced here. Until such re-analyses are performed, the results of many published ZDI studies of M dwarfs must be treated with caution.

\begin{acknowledgements}
The author thanks Dr. B. Edvardsson for sharing his collection of molecular line lists and partition functions. Support by the Swedish Research Council (grant agreement no. 2023-03667) and by the Swedish National Space Agency is gratefully acknowledged.
\end{acknowledgements}

\begin{appendix}

\section{Fast spectrum synthesis with {\sc FastSpec}}
\label{sec:fs}

In the course of the numerical calculations performed in this study, we recognised that existing spectrum synthesis tools, including {\sc Synmast}, are poorly suited to computing molecular opacities over large wavelength intervals. To address this limitation, we developed a new spectrum synthesis code, {\sc FastSpec}, which is specifically designed to handle extensive molecular line lists and is optimised for fast spectrum synthesis over large portions of stellar spectra.

{\sc FastSpec} is written in Fortran90, with many parts of the code parallelised using OpenMP. The code adopts continuum opacity routines from {\sc ATLAS12} \citep{kurucz:2005} and the equation-of-state solver from {\sc SME} \citep{piskunov:2017}\footnote{\url{https://github.com/MingjieJian/SMElib}}. The \ion{H}{i} bound-free and bound-bound opacities are based on the code by \citet{barklem:2003}\footnote{\url{https://github.com/barklem/hlinop}}, which has been converted to Fortran90 and optimised for parallel execution with OpenMP.

Calculations with {\sc FastSpec} proceed as follows. First, the code reads an input file specifying the wavelength setup and providing a set of files containing atomic and molecular line parameters. Atomic line data are taken from VALD \citep{ryabchikova:2015}\footnote{\url{https://vald.astro.uu.se}} without preselection (using the `extract all' option), while molecular line lists are obtained from the ExoMol database \citep{tennyson:2024}\footnote{\url{https://www.exomol.com}}. The line lists are stored in HDF5 format, which enables efficient compression and allows only those parts of the data relevant to a given wavelength interval to be read. After reading the input file, {\sc FastSpec} compiles a list of all relevant species from the available line lists and determines their concentrations and partition functions at all layers of the selected model atmosphere. Currently, the code supports {\sc MARCS} \citep{gustafsson:2008}, {\sc ATLAS} \citep{kurucz:2005}, and {\sc LLmodels} \citep{shulyak:2004} model atmospheres.

In the next step, {\sc FastSpec} computes the continuum opacity over the full wavelength range. The code then iterates over the HDF5 spectral line lists, and for each list the line opacities are summed over all relevant lines at each wavelength point. This part of the calculation is carefully optimised and results in substantial computational gains compared to other spectrum synthesis codes. These line opacity calculations comprise the following steps:
\begin{enumerate}[noitemsep, topsep=0.5pt]
\item Central line opacities are computed for individual spectral lines. Lines with a maximum line-to-continuum opacity ratio below $10^{-4}$ are excluded from further calculations. This line inclusion threshold is identical to that used by {\sc SME} and {\sc Turbospectrum} \citep{gerber:2023}\footnote{\url{https://github.com/bertrandplez/Turbospectrum_NLTE}}.
\item For the retained lines, the central opacity coefficients together with the corresponding Doppler and Lorentzian broadening parameters are stored for each spectral line and each atmospheric layer. These arrays are by far the most memory-intensive part of the calculation. Therefore, {\sc FastSpec} includes the option to perform line opacity calculations in chunks, depending on the available RAM.
\item Still treating each spectral line individually, {\sc FastSpec} determines the indices of the wavelength grid points to which the line wings contribute, based on the same opacity ratio criterion. After this step has been completed for all lines, the minimum and maximum indices of the lines contributing to each wavelength point are recorded.
\item Finally, a loop over all wavelengths and atmospheric layers is executed, in which all line opacities are summed. This loop exploits the precomputed line windows and line-list indices, thereby avoiding computationally expensive loops over all spectral lines.
\end{enumerate}
\smallskip
After completing the line opacity calculations, {\sc FastSpec} saves the line and continuum opacity files and optionally computes line and continuum intensity spectra for a predefined set of limb angles using the DELO-Bezier radiative transfer solver \citep{de-la-cruz-rodriguez:2013}.

In the context of this study, which focused on M-dwarf spectra at near-infrared wavelengths, {\sc FastSpec} was used to compute opacities with a velocity step of 0.5~\kms\ for a set of 17 isotopologues of 10 molecules (CH, OH, HF, NaH, CrH, FeH, CN, CO, TiO, and H$_2$O). The resulting molecular opacities were subsequently incorporated into the {\sc Synmast} polarised radiative transfer calculations described in Sect.~\ref{sec:prt}. 

In addition, we carried out test spectrum synthesis calculations over the 950--2500~nm wavelength range that included both atomic and molecular line opacity. For these calculations, the velocity step was reduced to 0.3~\kms, corresponding to a resolving power of $R\approx10^6$ and yielding about $10^6$ wavelength points. The full calculation using a $T_{\rm eff}=3500$~K, $\log g=5.0$ solar-metallicity {\sc MARCS} model atmosphere involved $135\times10^6$ spectral lines and required 235~s on an 18-core 2.3 GHz Intel Xeon W CPU, with the treatment of $131\times10^6$ H$_2$O lines dominating the computing time (222~s). Similar execution times were obtained on a 10-core M1 Pro CPU. For comparison, the TiO component of these calculations, comprising approximately $2\times10^6$ lines, was completed in only 3~s. Computing the same TiO spectrum required 1030~s with the latest version of {\sc Turbospectrum} and as much as 65~h with {\sc SME}, with both codes restricted to single-core execution. The superior computational performance of {\sc FastSpec} is maintained for purely atomic spectrum synthesis appropriate for hotter model atmospheres. For example, calculation of the full solar spectrum at seven limb angles over the 300--2500~nm interval at $R=10^6$, based on the $608\times10^3$ atomic lines from VALD, required only 12~s. This time increased to 37~s when calculations incorporated 292 hydrogen lines treated with the occupational probability formalism \citep{barklem:2003}.

\onecolumn

\section{ZDI studies of M dwarfs}
\label{sec:summary}

\begin{table*}[!h]
\caption{Summary of ZDI studies of M dwarfs that employed a filling-factor formalism for interpreting polarisation spectra.}
\label{tbl:zdi}
\centering
\begin{tabular}{l c l c c c c}
\hline\hline
Target & $\lambda$ range & Epoch & $B_{\rm max}$ (kG) & $f_{\rm V}$ & $B_{\rm max}/f_{\rm V}$ (kG) & References \\
\hline
OT Ser      & VIS & 2007--2008 & \textit{0.5}  & 0.05--0.10 & 5--10    & 1 \\
CE Boo      & VIS & 2008       & \textit{0.5}  & 0.05       & 10       & 1 \\
EV Lac      & VIS & 2006--2007 & \textit{2.0}  & 0.10--0.11 & 18--20   & 2 \\
            & NIR & 2020--2021 & 0.69--0.70    & 0.09--0.19 & 4--8     & 8 \\
YZ CMi      & VIS & 2007--2008 & \textit{2.5}  & 0.11       & 23       & 2 \\
AD Leo      & VIS & 2007--2008 & \textit{1.0}  & 0.14       & 7        & 2 \\
            & VIS & 2012--2016 & 0.95--1.18    & 0.07--0.13 & 9--14    & 5 \\
            & VIS & 2019       & 0.58          & 0.12       & 4.8      & 7 \\
            & NIR & 2019--2020 & 0.43--0.76    & 0.09--0.16 & 3.9--5.3 & 7 \\
            & NIR & 2023       & \textit{2.0}  & 0.15       & 13       & 10$^\dagger$ \\
EQ Peg A    & VIS & 2006       & \textit{1.0}  & 0.11       & 9        & 2 \\
GJ 51       & VIS & 2006--2008 & 3.86--5.02    & 0.12       & 32--42   & 3 \\
GJ 1245 B   & VIS & 2006--2008 & 0.22--0.47    & 0.06--0.10 & 2--6     & 3 \\
WX UMa      & VIS & 2006--2009 & 3.82--4.88    & 0.12       & 32--42   & 3 \\
DX Cnc      & VIS & 2008--2009 & 0.18--0.22    & 0.20       & 0.9--1.1 & 3 \\
GJ 1289     & VIS & 2016       & \textit{0.5}  & 0.15       & 3        & 4 \\
            & NIR & 2019--2022 & \textit{0.25} & 0.10       & 2.5      & 9 \\
Proxima Cen & VIS & 2017       & \textit{0.45} & 0.10       & 4.5      & 6 \\
DS Leo      & NIR & 2020--2022 & 0.04--0.12    & 0.06--0.10 & 0.6--1.2 & 8 \\
CN Leo      & NIR & 2019--2022 & 0.61--0.94  & 0.12--0.18 & 4.3--5.5 & 8 \\
GJ 905      & NIR & 2019--2022 & \textit{0.2}  & 0.10       & 2        & 9 \\
GJ 1151     & NIR & 2019--2022 & \textit{0.15} & 0.10       & 1.5      & 9 \\
GJ 1286     & NIR & 2020--2021 & \textit{0.2}  & 0.10       & 2        & 9 \\
GJ 617B     & NIR & 2020--2022 & \textit{0.1}  & 0.10       & 1        & 9 \\
GJ 408      & NIR & 2019--2022 & \textit{0.2}  & 0.10       & 2        & 9 \\
AU Mic      & NIR & 2019--2024 & \textit{1.5}  & 0.20       & 7.5      & 11 \\
            & NIR & 2023       & \textit{1.5}  & 0.20       & 7.5      & 10$^\dagger$ \\
YZ Cet & NIR & 2023 & 0.56 & 0.15 & 3.7 & 12 \\
            & NIR & 2023--2024 & 0.86--0.83 & 0.11 & 7.6--7.8 & 13 \\
\hline
\end{tabular}
\tablefoot{Italic values in the $B_{\rm max}$ column denote approximate estimates from the ZDI maps.
$^\dagger$Four Stokes parameter study applying the same filling factor to $V$ and $QU$ profile modelling.}
\tablebib{(1) \citet{donati:2008c}; (2) \citet{morin:2008}; (3) \citet{morin:2010}; (4) \citet{moutou:2017}; (5) \citet{lavail:2018}; (6) \citet{klein:2021a}; (7) \citet{bellotti:2023a}; (8) \citet{bellotti:2024}; (9) \citet{lehmann:2024}; (10) \citet{donati:2025a}; (11) \citet{donati:2025}; (12) \citet{pineda:2025}; (13) \citet{biswas:2025}.}
\end{table*}

\end{appendix}

\end{document}